\documentclass[twocolumn]{aa}

\usepackage{graphicx}
\usepackage{txfonts}
\usepackage[dvipsnames]{xcolor}
\usepackage{soul}
\usepackage{hyperref}
\usepackage[capitalise]{cleveref}
\usepackage{siunitx}
\usepackage{orcidlink}

\def \parantheses  #1{\left(#1\right)} 
\def \braces       #1{\left\{#1\right\}} 

\hypersetup{
    colorlinks,
    linkcolor={red!50!black},
    citecolor={blue!50!black},
    urlcolor={blue!80!black}
}

\DeclareSIUnit \year {yr}
\DeclareSIUnit \parsec {pc}
\DeclareSIUnit \eV {eV}
\DeclareSIUnit \Msun {M_\odot}

\begin{document}
   \title{Modelling the track of the GD-1 stellar stream inside a host with a fermionic dark matter core-halo distribution}

   \author{Martín F. Mestre\inst{1,2}\fnmsep\thanks{
         \email{mmestre@fcaglp.unlp.edu.ar}}\orcidlink{0009-0001-9329-5260}
      \and
      Carlos R. Argüelles\inst{1,3}\orcidlink{0000-0002-5862-8840}
      \and
      Daniel D. Carpintero\inst{1,2}
      \and
      Valentina Crespi\inst{2}
      \and
      Andreas Krut\inst{3}
   }

   \institute{Instituto de Astrof{\'i}sica de La Plata (CONICET-UNLP), Paseo del
               Bosque S/N, La Plata (1900), Buenos Aires, Argentina\\
               \and
              Facultad de Ciencias Astron{\'o}micas y Geof{\'i}sicas de La Plata (UNLP),
              Paseo del Bosque S/N, La Plata (1900), Buenos Aires, Argentina\\
              \and
              ICRANet, Piazza della Repubblica 10, 65122 Pescara, Italy\\
             }

   \date{Accepted on 23/04/2024 by A$\&$A}


\abstract{
    Traditional studies on stellar streams typically involve phenomenological $\Lambda$CDM halos or \textit{ad hoc} dark matter (DM) profiles with different degrees of triaxiality, which preclude to gain insights into the nature and mass of the DM particles. Recently, a Maximum Entropy Principle of halo formation has been applied to provide a DM halo model which incorporates the fermionic (quantum) nature of the particles, while leading to DM profiles which depend on the fermion mass. Such profiles develop a more general \textit{dense core} - \textit{diluted halo} morphology able to explain the Galactic rotation curve, while the degenerate fermion core can mimic the central massive black hole (BH).
}
{
    We attempt to model the GD-1 stellar stream using a spherical \textit{core}-\textit{halo} DM distribution for the host, which, at the same time, explains the dynamics of the S-cluster stars through its degenerate fermion-core with no central BH.
}
{
    We used two optimization algorithms in order to fit both the initial conditions of the stream orbit and the fermionic model. We modelled the baryonic potential with a bulge and two disks (thin and thick) with fixed parameters according to the recent literature. The stream observables are 5D phase-space data from the Gaia DR2 survey.
}
{
    We were able to find good fits for both the GD-1 stream and the S-stars for a family of fermionic \textit{core}-\textit{halo} profiles parameterized by the fermion mass. Particle masses are constrained in the range $\SI{56}{\kilo\eV\per c^2}$ -with corresponding DM core of $\sim \num{E3}$ Schwarzschild radius- all the way to $\SI{360}{\kilo\eV\per c^2}$ corresponding to the most compact core of $5$ Schwarzschild radius prior to the gravitational collapse into a BH of about $\SI{4E6}{\Msun}$.
}
{
    This work provides evidence that the fermionic profile is a reliable model for both the massive central object and the DM of the Galaxy. Remarkably, this model predicts a total MW mass of $\SI{2.3E11}{\Msun}$  which is in agreement with recent mass estimates obtained from Gaia DR3 rotation curves (Gaia RC). In summary, with one single fermionic model for the DM distribution of the MW, we obtain a good fit in three totally different distance scales of the Galaxy: $\sim 10^{-6}$ kpc (central, S-stars), $\sim14$ kpc~(mid, GD-1) and $\sim 30$ kpc~(boundary, Gaia RC mass estimate).
}

   \titlerunning{The GD-1 stream inside a fermionic dark matter halo}

   \maketitle

\section{Introduction}
The gravitational interaction between a galaxy and a satellite (a dwarf galaxy or a globular cluster) modifies both systems: the host galaxy strips stars from the satellite at a rate
that depends on their density profiles and on the orbit,
while the density profile of the host suffers a reorganization of its matter in the vicinity of the satellite's orbit, developing a wake due to the dynamical friction.
The ensemble of stars tidally stripped from the satellite constitute a so-called stellar stream (also known as tidal stream); examples of them have been detected in the Milky Way (MW), Andromeda and the Local Volume~\citep{Mart_Delgado_2010}.

At present, stellar streams constitute one of the main MW observables related to the dynamics, together with other baryonic observables like the Galaxy rotation curve, the radial surface density profile of the disk, and the vertical density profile of the disk at the solar radius.
Tidal streams probe the acceleration field produced by the Galaxy~\citep{1999ApJ...512L.109J,1999A&A...348L..49Z,2009ApJ...703L..67L,2013MNRAS.436.2386L,2016ASSL..420..169J,Ibata_2016,2017ApJ...842..120I,2017A&A...603A..65T,2021MNRAS.502.4170R} as well as its formation history according to ~\citet{1999Natur.402...53H}, \citet{2020ARA&A..58..205H}, \citet{2022A&A...666A..64R}, and \citet{2023arXiv230708730C}, among others.
Indeed, a stellar stream could be the closest realization of a galactic orbit that can be observed in Nature.
Nevertheless, the larger the progenitor, the more the discrepancy between its orbit and the stream phase-space configuration. In fact, the taxonomy of streams is very rich \citep{2015MNRAS.450..575A} because of the different gravitational configurations that can take place, going from almost one-dimensional streams whose progenitor is a small globular cluster, to wide shells produced by the partially radial sinking of a large progenitor into the MW centre. Besides, it is theoretically possible to have an accreted stream whose progenitor is a globular cluster orbiting a MW's satellite galaxy, giving rise to stellar cocoons around the
stream track \citep{2018ApJ...861...69C,2019ApJ...881..106M,2021MNRAS.501..179M,2021ApJ...911L..32G,2022MNRAS.511.2339Q}.
Moreover, it is also theoretically possible to have perturbations of the streams due to dark matter sub-halos, forming off-track features like the detected spur~\citep{Price-Whelan_2018} in
the GD-1 stream~\citep{Grillmair_2006}.

Stream data together with other measurements from baryonic tracers can help in making claims about an unknown aspect of the gravitational field: whether to model it with a dark matter (DM) halo component, or a Modified Newtonian Dynamics (MoND) theory is needed.
For example, within the DM paradigm, spatial densities might have different profiles
(e.g. spherical, axisymmetric, triaxial) depending on the number of conserved components of the angular momentum, which also influences the stream properties \citep{2013ApJ...773L...4V,2016MNRAS.455.1079P,2020MNRAS.492.4398M}.
In particular, \citet{2019MNRAS.486.2995M} have constrained the MW dark matter halo shape using Gaia DR2 data of the GD-1 stream, assuming an axisymmetric generalization of the NFW profile and obtaining a flattening halo parameter of $q=0.82^{+0.25}_{-0.13}$, thus compatible with spherical symmetry.
Moreover, \cite{2023MNRAS.524.2124P} were able to measure the oblateness in an axisymmetric generalization of the NFW profile using three stellar streams: NGC~3201, M68 and Palomar~5, obtaining
consistency with a spherical halo.

In addition to the above traditional halo models, which arise from $\Lambda$CDM cosmologies or from various ad hoc symmetry considerations, there are DM profiles that take into account the quantum nature of the DM candidate: see e.g. \cite{2014NatPh..10..496S} for bosonic profiles (composed of axion-like particles), and \cite{2015MNRAS.451..622R,2015PhRvD..92l3527C,2021MNRAS.502.4227A} for fermionic profiles (typically composed of sterile neutrinos). A relevant aspect of such kind of profiles is the source of quantum pressure acting in the innermost regions of the halos: while in the boson case the profiles develop a highly dense soliton, in the fermion case the profiles develop a degenerate compact core surrounded by a more dilute halo which is self-bounded in radius. In this work we focus on the latter. Such a \emph{dense~core~--~diluted~halo} fermionic DM profile is obtained from the fully relativistic Ruffini-Argüelles-Rueda (RAR) model, which was successfully applied to different galaxy types \citep{2019PDU....24..278A,2023ApJ...945....1K} including the Milky Way from Sagittarius~A* till the entire halo \citep{arguelles_novel_2018,2020A&A...641A..34B,2021MNRAS.505L..64B,2022MNRAS.511L..35A}.
This kind of model built in terms of a Fermi-Dirac-like phase space distribution function (including for central degeneracy and cut-off in the particle energy), is also known as the relativistic fermionic King (RFK) model \citep{2022PhRvD.106d3538C}.

In this paper we model the 5D track of the GD-1 stellar stream inside a MW with a fermionic dark matter core-halo distribution. At the same time, we aim to explain the dynamics of the S-cluster stars through the high density fermion-core and without assuming a central BH. Finally we compare the virial mass of the Galaxy as predicted by the fermionic model with that obtained from recent Gaia DR3 rotation curve data. In Sec.~\ref{sec:methodology} we explain the methodology; in Sec.~\ref{sec:results} we present the best-fit results, and in Sec.~\ref{sec:conclusions} we give the conclusions.

\section{Methodology}
\label{sec:methodology}
In this section we explain the observables and methods used in this research.
The exact pipeline applied in order to obtain the results and plots of this paper
can be found at the following GitHub repository:
\url{https://github.com/martinmestre/stream-fit/blob/main/pipeline_paper/}.

\subsection{Observables and assumed measurements}
\label{sec:observables}

The main observables that we have used were computed with the polynomial fits found by
\citet{Ibata_2020} for the \mbox{GD-1} stream using astrometric (Gaia DR2) and high-precision spectroscopic datasets, together with the analysis of the {\sc streamfinder} algorithm.
The polynomials are the following:
\begin{align}
   \label{Ibata_polyn}
   \phi_2  &= 0.008367\phi_1^3-0.05332\phi_1^2-0.07739\phi_1-0.02007, \\
   \tilde{\mu}_\alpha &= 3.794\phi_1^3+9.467\phi_1^2+1.615\phi_1-7.844,\\
   \mu_\delta &= -1.225\phi_1^3+8.313\phi_1^2+18.68\phi_1-3.95,\\
   v_h &=  90.68\phi_1^3+204.5\phi_1^2-254.2\phi_1-261.5,
\end{align}
with $\phi_1$ and $\phi_2$ in radians, $\tilde{\mu}_\alpha=\mu_\alpha \cos \delta$ and $\mu_\delta$ in mas yr$^{-1}$ and $v_h$ in km s$^{-1}$. These quantities correspond to the longitude and latitude in the GD-1 celestial frame of reference~\citep{Koposov_2010}; the proper motion in right ascension and declination, and the heliocentric radial velocity respectively.
The domain of the polynomials is limited to $-90^\circ <\phi_1<10^\circ$.
To obtain our observable data set we sampled the domain with 100 equidistant points ($\phi_1^{(i)},\, i=1,\dots,100$)
and evaluated the polynomials at those points, thus obtaining the sets $\phi_1^{(i)}$, $\phi_2^{(i)}$, $\tilde{\mu}_\alpha^{(i)}$, $\mu_\delta^{(i)}$ and $v_h^{(i)}$. Note that we have not included, among the observables in Eq.~(\ref{Ibata_polyn}), the photometric heliocentric distance $D$, justified by the posterior analysis performed in Sec.~\ref{sec:fitting}.
In one of the experiments we will consider an `observable' of a different nature,
the core mass ($M_{\rm{core}}$), which is defined as the mass enclosed at the radius when the circular velocity reaches its first maximum.
The constraint for the mass of the core of the distribution is assumed to be $M_{\rm core } = \SI{3.5E6}{\Msun}$ in agreement with~\citet{2020A&A...641A..34B,2021MNRAS.505L..64B} and \citet{2022MNRAS.511L..35A}.
The core radius does not include all the mass inside the innermost S-star pericenter here considered (S2), because the first maximum of the circular velocity corresponds to a shorter distance in which the core density-region is still falling. Indeed, in Sec. \ref{sec:results} we obtain a DM mass inside the S2 star pericenter of $M(r_{peri-\rm{S2}}) = 4.03\times 10^6 M_\odot$, in excellent agreement with the Schwarzschild BH mass constraints of $4.1\times 10^6 M_\odot$ and $3.97\times 10^6 M_\odot$ obtained from the same S2 star in \cite{2018A&A...615L..15G} and \cite{2019Sci...365..664D} respectively.
The parameters assumed in this paper are the
Galactocentric distance of the Sun $R_\odot=\SI{8.122}{\kilo\parsec}$ \citep{2018A&A...615L..15G} and the Sun's peculiar
velocity\footnote{
We adopted a Cartesian reference frame $(X, Y, Z)$ with
corresponding velocities $(U, V, W)$ in which the $X$ and $U$ axes
point from the Galactic center towards the opposite direction
of the Sun; $Y$ and $V$ point in the direction of the Galactic rotation at the
location of the Sun; and $Z$ and $W$ point
towards the North Galactic Pole. Note that this is the same right-handed reference system adopted by Astropy.} $\vec{v}_{\odot p} = (\SI{11.1}{\kilo\metre\per\second}, \SI{12.24}{\kilo\metre\per\second}, \SI{7.25}{\kilo\metre\per\second})$ \citep{Shonrich}.

\subsection{The fermionic DM halo model}
\label{sec:fermionicDM}
Our fermionic DM model is a spherical and isotropic distribution of fermions at finite temperature in hydrostatic equilibrium, subject to the laws of General Relativity (GR), i.e. the T.O.V. equation complemented with the Tolman and Klein thermodynamics equilibrium conditions and the particle energy conservation along a geodesic, as defined in~\cite{arguelles_novel_2018} while using a notation from~\cite{2020EPJP..135..290C}, detailed below.

We start with a spherically symmetric metric defined as:
\begin{equation}
    \label{metric}
    \mathrm{d}s^2 = g_{00}(r)\mathrm{d}t^2 + g_{11}(r)\mathrm{d}r^2 -r^2\mathrm{d}\vartheta^2 -r^2\sin\vartheta \mathrm{d}\varphi^2,
\end{equation}
with $g_{00}(r) = \mathrm{e}^{\nu(r)}c^2$ and $g_{11}(r) = -\mathrm{e}^{\lambda(r)}$,
where $c$ is the speed of light, $t$ stands for the time, ($r$, $\vartheta$, $\varphi$) are spherical coordinates and $\nu$ and $\lambda$ are metric exponents whose radial dependence will be computed below.

Our first differential equation is that of the mass versus radius for a spherical system of density $\rho$:
 \begin{equation}
    \label{mass_def}
    \frac{\mathrm{d}M}{\mathrm{d}r} = 4\pi r^2 \rho(r),
 \end{equation}
from which we obtain the enclosed mass $M(r)$ at a given radius $r$ by simple integration:
\begin{equation}
 M(r)=4\pi\int_0^r r'^2 \rho(r') \mathrm{d}r'.
\end{equation}
From the Einstein equations, the relation between $M(r)$ and the metric exponent $\lambda$ can be found:
\begin{equation}
    \mathrm{e}^{-\lambda(r)}=1-\frac{2G}{c^2}\frac{M(r)}{r},
\end{equation}
where $G$ is the gravitational constant.

The following version of the T.O.V. equation for the metric exponent $\nu$ can be deduced:
\begin{align}
   \label{tov}
    \frac{\mathrm{d}\nu}{\mathrm{d}r}= \frac{1+ \frac{\displaystyle 4\pi r^3 P(r)}{\displaystyle M(r)c^2}}{r\left(\frac{\displaystyle rc^2}{\displaystyle 2GM(r)} -1\right)},
\end{align}
where $P$ is the pressure.
Both quantities $\rho$ and $P$ are defined as the following integrals over momentum space:
\begin{align}
     \label{dens_press}
      \rho(r)&=\int \frac{E(p)}{c^2}~f(r,p)\mathrm{d}\boldsymbol{p},\\
      P(r)&=\frac{1}{3}\int p\frac{dE(p)}{dp}~f(r,p)\mathrm{d}\boldsymbol{p}=
                \frac{1}{3}\int \frac{p^2c^2}{E(p)}~f(r,p)\mathrm{d}\boldsymbol{p},
\end{align}
where $\boldsymbol{p}$ is the spatial momentum  vector, $p$ is its norm, $E(p)=\sqrt{p^2c^2+m^2c^4}$ is the total energy of a particle of mass $m$, and $f$ is the phase-space distribution of the system, given by a Fermi-Dirac distribution with energy cut-off. This number density $f$ can be obtained from a maximum entropy principle computed from a kinetic theory that includes self-gravity and violent relaxation, as shown in \citealp{2004PhyA..332...89C}~\citep[for a review see also][]{2022PhyA..60628089C}, and was recently applied to a vast sample of disk galaxies in \cite{2023ApJ...945....1K}. It can be expressed as
\begin{equation}
f(r,p)=
    \frac{g}{h^3}
      \frac{1-\mathrm{e}^{[E(p)-E_\mathrm{c}(r)]/kT(r)}}
      {1+\mathrm{e}^{[E(p)-\mu(r)]/kT(r)}}\quad\mathrm{if}\quad E(p) \leq E_\mathrm{c}(r),\\
\end{equation}
and $f(r,p)=0$ otherwise,
where $k$ is the Boltzmann constant, $h$ is the Planck constant, $g=2s+1$ is the spin multiplicity of quantum states, with $s=1/2$, and the following local quantities are used: chemical potential $\mu(r)$, cut-off energy $E_\mathrm{c}(r)$,  and effective temperature $T(r)$.
The coefficient $g/h^3$ is the maximum accessible value of the distribution function fixed by the Pauli exclusion principle.

The above equations are complemented with two thermodynamic equilibrium conditions given
by~\citet{PhysRev.35.904} and~\citet{RevModPhys.21.531}, together with the condition of energy conservation
along the geodesic given in~\citet{1989A&A...221....4M}:
\begin{equation}
    \label{tke}
   \frac{1}{T}\frac{\mathrm{d}T}{\mathrm{d}r}=\frac{1}{\mu}\frac{\mathrm{d}\mu}{\mathrm{d}r}=
   \frac{1}{E_\mathrm{c}}\frac{\mathrm{d} E_\mathrm{c}}{\mathrm{d}r}=-\frac{1}{2}\frac{\mathrm{d}\nu}{\mathrm{d}r}.
\end{equation}

Thus, we have built a system of five differential equations given
by Eqs.~(\ref{mass_def}), (\ref{tov}), and (\ref{tke}), with initial conditions
at the centre of the distribution $M(0)=0$, $\nu(0)=0$, $T(0)=T_0$,
$\mu(0)=\mu_0$ and $E_c(0)=E_{c0}$.
Note that the differential equations do not depend on $\nu$ but on its radial
derivative, so the system can be solved starting with an arbitrary initial value $\nu(0)=0$, adding afterwards a finite value, namely $\nu_0$, in such a way that the solution satisfies a condition of continuity with the Schwarzschild metric at the border of the fermionic distribution. In Appendix~\ref{app:numerical} we explain how the system of equations was solved numerically. Following~\cite{arguelles_novel_2018}, throughout this paper we use adimensional versions of the initial conditions:
\begin{align}
    \theta_0 &= \frac{\mu_0 - mc^2}{kT_0},\nonumber\\
    W_0 &= \frac{E_{c0} - mc^2}{kT_0},\nonumber\\
    \beta_0 &= \frac{kT_0}{mc^2}.\label{rar_params}
\end{align}
Note that we have subtracted the rest-energy in the first two cases.

Although our fermionic system is univocally determined by the four free parameters $m$, $\theta_0$, $W_0$ and $\beta_0$, in some convenient situations we will use $\omega_0=W_0-\theta_0$ instead of $W_0$.

\subsection{Milky Way and stream models}
\label{sec:MW_stream_models}

We modelled our Galaxy by combining the fermionic dark halo described above --- whose parameters will be determined in this work --- with a fixed baryonic component identical to the one in Model I of \citet{2017A&A...598A..66P}. We name this full Galaxy model as Fermionic-MW.

Besides the above mentioned model, and for qualitative comparative purposes, we will make use of the Galactic model fitted by \citet{2019MNRAS.486.2995M}, which is the {\texttt{MWPotential2014}} model with an axisymmetric NFW profile. This model was evaluated by means of the {\it Galpy} package~\citep{2015ApJS..216...29B}. A robust statistical comparison between the RAR and other models will be performed in a future paper whereas here we intend to verify that our GD-1 fit is in agreement with the latest fit performed in the literature. This fitted model uses a circular velocity at the position of the Sun $v_\mathrm{c}(R_\odot)=244 \pm 4$ km s$^{-1}$ and a $z-$flattening of the DM density distribution  $q_\rho=0.82^{+0.25}_{-0.13}$. We name this Galaxy model as NFW-MW.

GD-1 is a dynamically cold stream, with its stars keeping a large degree of correlation, though
its progenitor has not been detected with certainty~\citep{10.1093/mnras/sty677,Price-Whelan_2018,10.1093/mnras/sty1338}.
But its almost one-dimensional distribution in phase-space could be well approximated with the orbit of a theoretical progenitor, as previoulsy done by~\cite{2019MNRAS.486.2995M,Price-Whelan_2018} and \cite{2010ApJ...712..260K}.

The initial conditions of the orbit were given in the spherical equatorial coordinates of the ICRS frame: right ascension $\alpha$, declination $\delta$, $D$, $\tilde{\mu}_\alpha$, $\mu_\delta$, and $v_h$. The code uses the {\it Astropy} ecosystem~\citep{astropy:2022, astropy:2018, astropy:2013} in order to transform these initial conditions to Galactocentric coordinates assuming a Galactocentric reference frame with the Sun at the position $\vec{x}_\odot=(-R_\odot,0,0)$ and a Sun's velocity given by the sum of the circular velocity at the position of the Sun and the Sun's peculiar velocity: $\vec{v}_\odot = \vec{v}_{\odot p} + (0, v_{\rm{c}}(R_\odot), 0)$. The circular velocity depends on the model and position and is given by \begin{equation}
\label{circ_vel}
    \begin{split}
       v_{\rm{c}}^2(R_\odot) &= R_\odot|\nabla \Phi(\vec{x})|_{\vec{x}=\vec{x}_\odot}\\
       &= R_\odot|\nabla \Phi_{\rm{B}}(\vec{x})|_{\vec{x}=\vec{x}_\odot} + G \frac{M_{\rm{DM}}(R_\odot)}{R_\odot}
    \end{split}
\end{equation} where $\Phi$ is the total potential, $\Phi_B$ is the potential generated by the three baryonic components and $M_{\rm{DM}}$ is the enclosed fermionic DM mass. In the last term, we have used the spherical symmetry of the DM distribution in order to relate the acceleration with the enclosed mass. For the NFW-MW model, the circular velocity was computed using the gradient of the total potential, i.e. first line of Eq.~(\ref{circ_vel}), performed with the {\it Galpy} code.

We integrated the orbit forwards and backwards in time during a time interval of $\Delta t=\SI{0.2}{\giga\year}$, starting in both cases from a given initial condition for the progenitor. In the next sections we will explain how these initial conditions were chosen for some simulations, and fitted for others. The integrator used was a Runge-Kutta of order eight (\texttt{DOP853} called from {\it SciPy}'s \texttt{solve\_ivp} function) with relative and absolute tolerance parameters given by \texttt{rtol}$=\num{5E-14}$ and \texttt{atol}$=\num{0.5E-14}$, respectively.

The resulting orbit was successively transformed to the ICRS and then to the GD-1 frames of coordinates. For the latter we used the \texttt{GD1Koposov10} class defined in the \texttt{Gala} package~\citep{gala,adrian_price_whelan_2020_4159870} which uses the transformation matrix defined by \citet{Koposov_2010}.
After these transformations we obtained the orbit expressed in the observable variables $\phi_1$, $\phi_2$, $\tilde{\mu}_\alpha$, $\mu_\delta$, $v_h$. Finally, we computed these variables at the points $\phi_1^{(i)}$ by interpolation. These values, together with the observed GD-1 data defined in Sec.~\ref{sec:observables}, were used to evaluate the following stream function:
\begin{align}
   \label{chi2:stream}
   \chi^2_{\rm{stream}} &= \chi^2_{\phi_2} + \chi^2_{\tilde{\mu}_\alpha}+ \chi^2_{\mu_\delta}+\chi^2_{v_h}\\
   \label{chi2:component}
   \chi^2_{\eta} &= \frac{1}{\sigma_\eta^2}\sum_{i=1}^{100} \parantheses{\eta^{(i)}-\eta(\phi_1^{(i)})}^2
\end{align}
where
$\eta\in\braces{\phi_2, v_h, \tilde{\mu}_\alpha, \mu_\delta}$, and $\sigma_\eta$ are the corresponding dispersions of the stream data points estimated by inspection of Figs.~1 and 3 of~\citet{Ibata_2020}:
$\sigma_{\phi_2}=0\fdg 5$,
$\sigma_{v_h} = 10$~km s$^{-1}$, and
$\sigma_{\tilde{\mu}_\alpha}= \sigma_{\mu_\delta}= 2$~mas yr$^{-1}$.
Thus, $\chi^2_{\rm{stream}}$ measures the departure of the model from the observed stream.

For some fits we will also consider the departure of the model from a dark mass constraint in the core
of the distribution:
\begin{equation}
   \label{chi2:core}
\chi^2_{\rm{core}}=\frac{(m_\mathrm{c}-M_{\rm{core}})^2}{\sigma_m^2},
\end{equation}
where the value of $M_{\rm{core}}$ was defined in Sec.~\ref{sec:observables}, $m_\mathrm{c}$ is the core mass of the model (i.e. variable), and
$\sigma_m$ was fixed at $0.01 M_{\rm{core}}$.
For these fits we will then use the following compound function:
\begin{equation}
   \label{chi2:full}
\chi^2_{\rm{full}}=\chi^2_{\rm{stream}}+\chi^2_{\rm{core}}.
\end{equation}
We note that in order to compute the estimated core mass for each model, $m_\mathrm{c}$ in \cref{chi2:core}, we need to calculate the first local maximum of the circular velocity in GR, the latter compactly expressed as:
\begin{equation}
   \label{v_circ_dm}
   V_{\rm{circ, DM}}(r)= c\left(\frac{r}{2}\frac{g_{00}'(r)}{g_{00}(r)}\right)^{1/2} =
                        c\left(\frac{r}{2}\frac{\mathrm{d}\nu(r)}{\mathrm{d}r}\right)^{1/2},
\end{equation}
where the velocity has components $V^i=dx^i/dt~(i=1,2,3)$ measured in a local frame that is fixed in space at a distance $r$ from the Galaxy centre. From the T.O.V.~equation~(\ref{tov}) it is possible to show that far from the core this relativistic formula for the velocity tends to the classical law, i.e. $\sqrt{G~M(r)/r}$.

\subsection{Optimization algorithms}
\label{sec:optimization}

Our goal is to fit our Fermionic-MW model by minimizing the full $\chi^2$ function given by \cref{chi2:full} (when fitting the NFW-MW model we will use only the function given by \cref{chi2:stream}). To this end we will use two optimization algorithms that belong to the family of Black Box algorithms which are very performant when the function to be minimized presents many relative minimums or the landscape is complex like in our case.

One is an implementation of a differential evolution algorithm, which is a metaheuristic algorithm that finds the solution of an optimization problem by iteratively updating generations of candidate solutions until a certain tolerance is met. Generally, a few best candidates from each generation survive in order to create the descendants, i.e. the next generation, by making stochastic combinations from them. We have used {\it SciPy}'s implementation of this algorithm, called \texttt{optimize.differential\_evolution} algorithm, with metaparemeters given by \texttt{strategy}=\texttt{"best2bin"}, \texttt{maxiter}=200, \texttt{popsize}=200, \texttt{tol}=$\num{5E-8}$ and \texttt{atol}=0, unless otherwise stated. This method can be run in parallel with shared memory.

The other algorithm is an implementation of the Mesh Adaptive Direct Search (MADS) algorithm called
\texttt{NOMAD}~\citep{audet2021nomad}. \cite{MADS_2006} give a detailed explanation of the method. The {\sc Julia}~\citep{bezanson2017julia}
wrapper of this algorithm, \href{https://bbopt.github.io/NOMAD.jl/stable/}{\it NOMAD.jl},
was used through the package \href{https://docs.sciml.ai/Optimization/stable/}{\it Optimization.jl}.
We used default values of all the metaparameters except for \texttt{maxiters}=700.

\section{Results}
\label{sec:results}

\subsection{Fitting the Fermionic-MW model}
\label{sec:fitting}
We fitted both the Fermionic-MW model parameters and the initial conditions (IC) of the orbit of the progenitor through four steps that consisted in: (i) obtaining an order zero value of the orbit IC using the
NFW-MW model, (ii) fitting the Fermionic-MW parameters with fixed IC values, (iii) polishing the IC values with fixed Fermionic-MW potential and (iv) polishing the Fermionic-MW parametes with fixed IC values.

As a first step we searched for a provisional but good enough set of IC that can reproduce the orbit in our
Fermionic-MW model. To this end, we fitted the initial conditions in the fixed NFW-MW model by using the $\chi^2_{\rm{stream}}$ function as defined in \cref{chi2:stream}
and the differential evolution algorithm specified in Sec.~\ref{sec:optimization}. Since the optimization algorithm
needs bounds for the variables, we used boxes centered near the midpoint
of the observable data, i.e. $\eta^{(51)}$, that corresponds to $\alpha=149\fdg 24$, $\delta=36\fdg 61$, $\tilde{\mu}_\alpha=-5.70$ mas yr$^{-1}$, $\mu_\delta=-12.48$ mas yr$^{-1}$ and $v_h=-18.81$ km s$^{-1}$
plus $D=7.69$ kpc,\footnote{This value corresponds to evaluating $D(\phi_1^{(51)})$ according to Eqn.~(\ref{phot_dist}).} and with sides of length equal to the absolute values of the variables at their centers (except for $\alpha$, where we have used a side of length $\alpha/5$).

\begin{figure}
   \centering
   \includegraphics[width=\hsize]{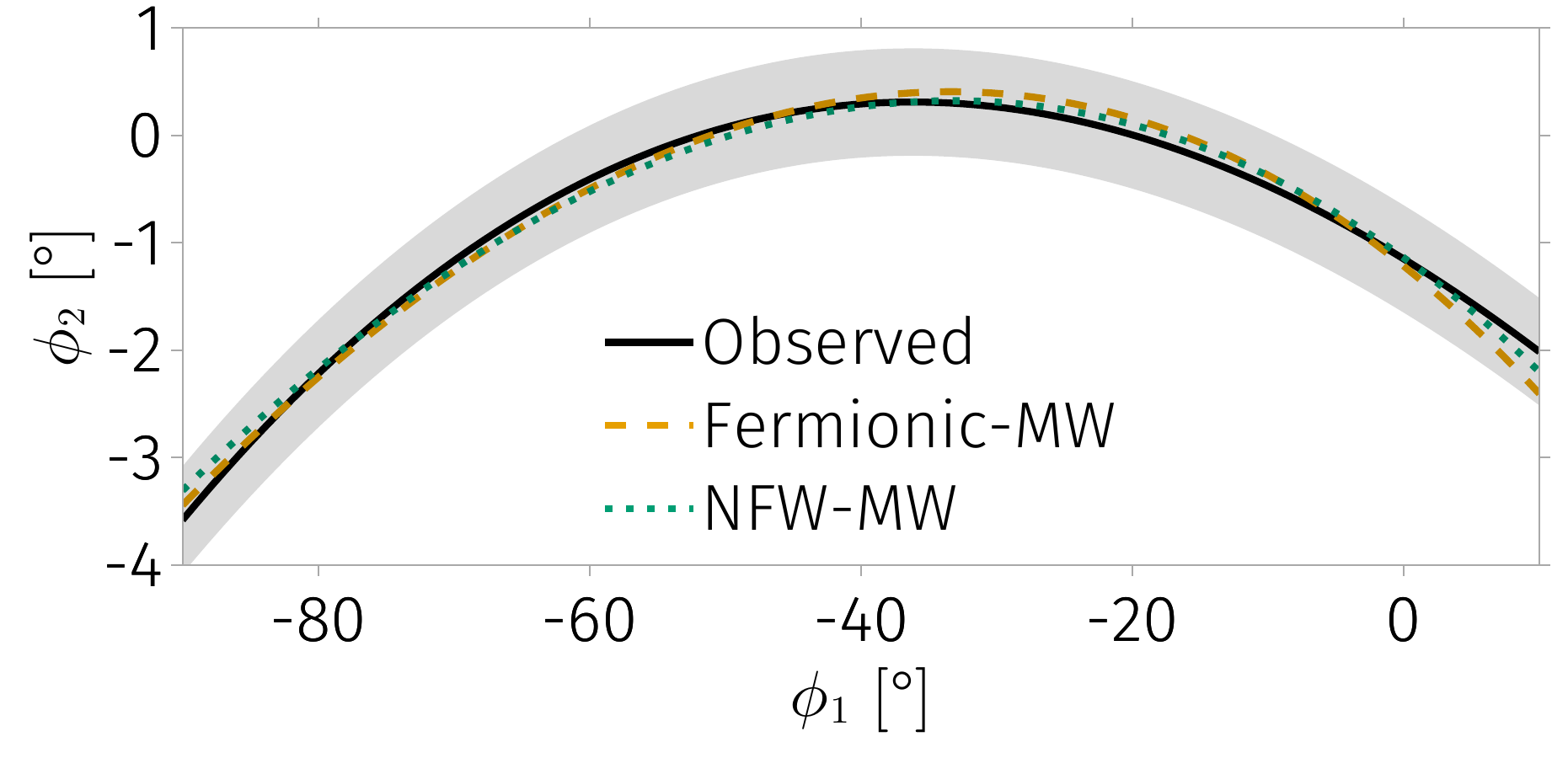}
   \includegraphics[width=\hsize]{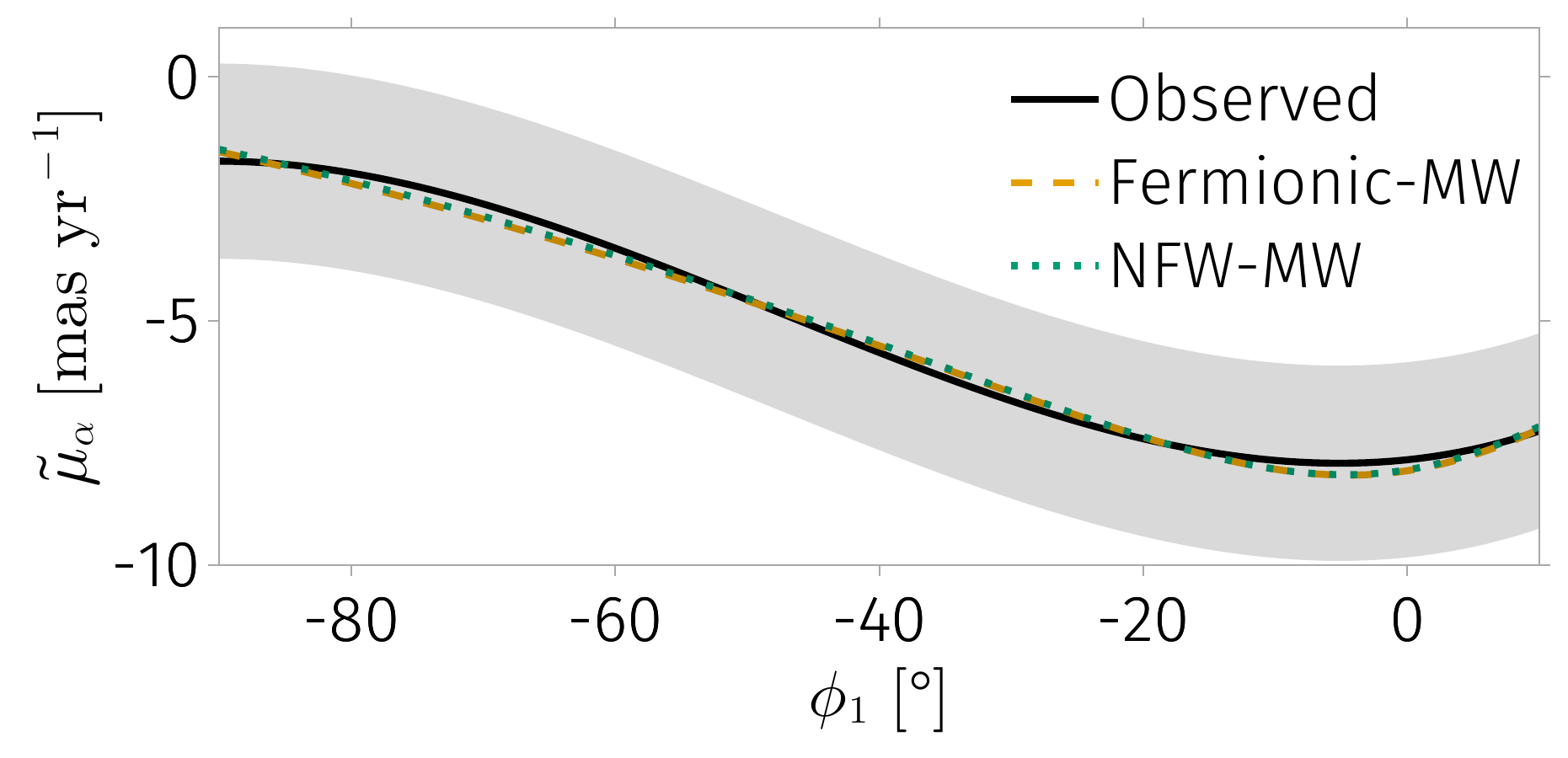}
   \includegraphics[width=\hsize]{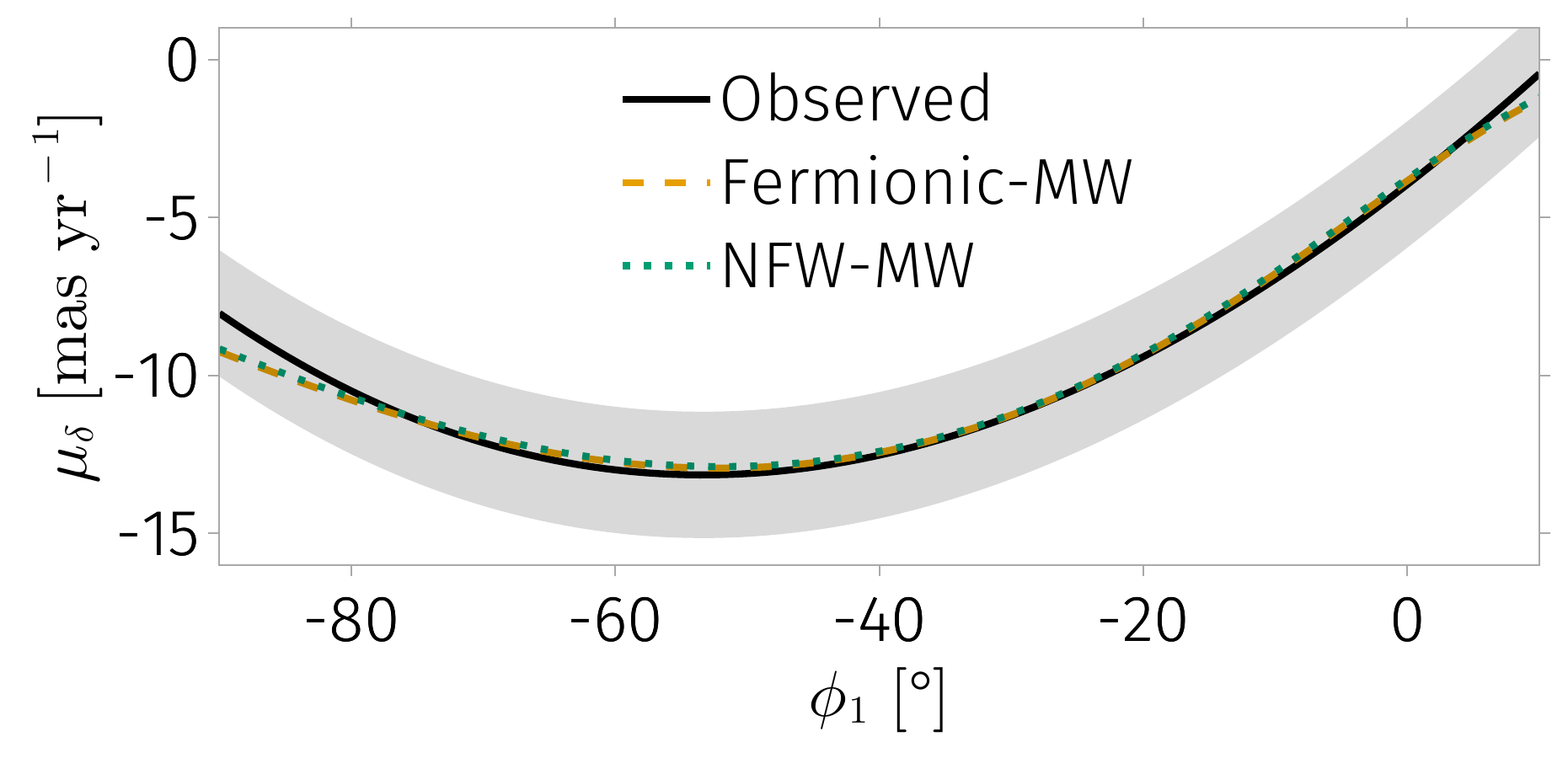}
   \includegraphics[width=\hsize]{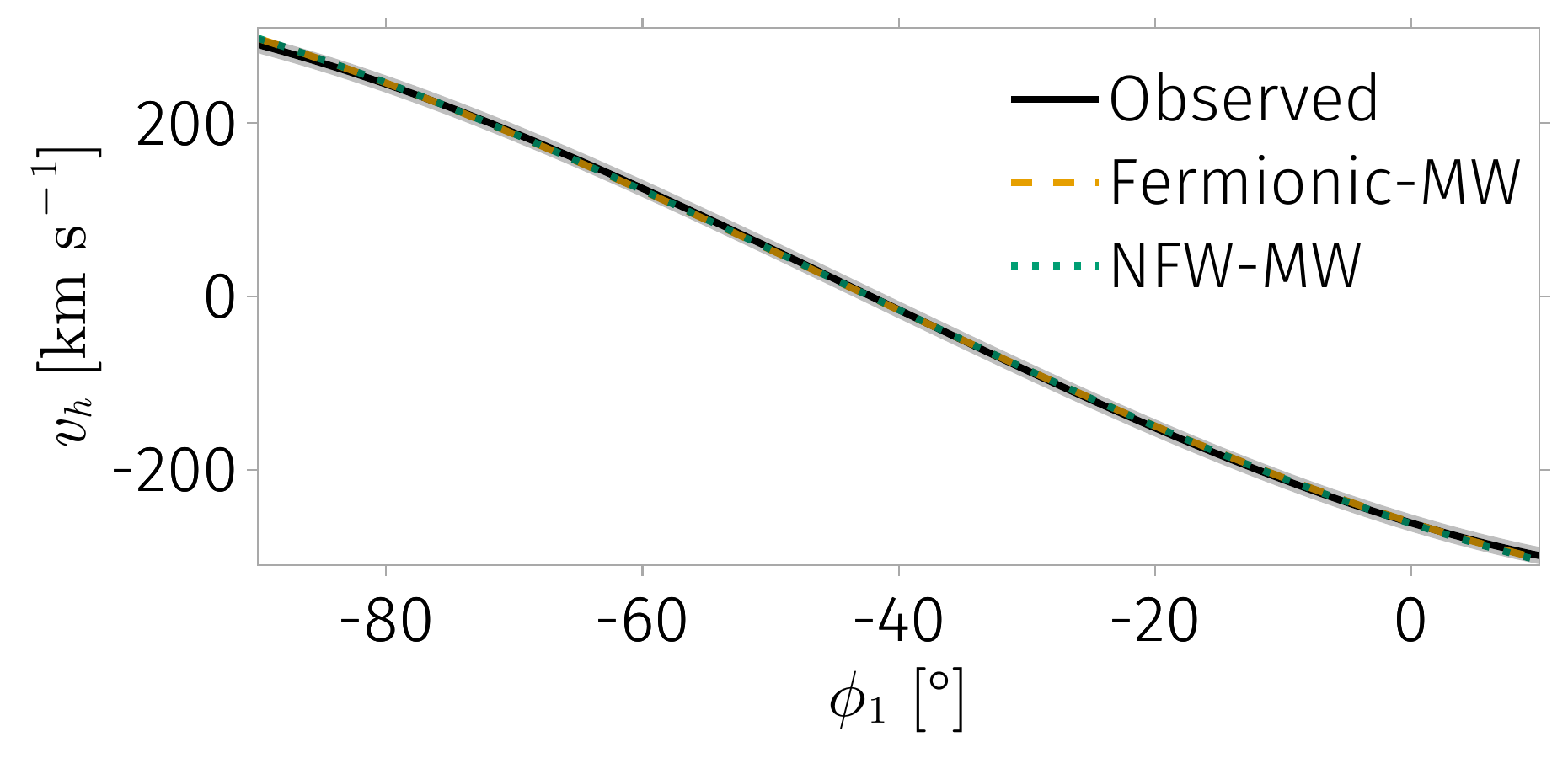}
   \caption{Stream fits in observable space: sky position (top: $\phi_2$), proper motions (middle: $\tilde{\mu}_\alpha$, $\mu_\delta$) and heliocentric velocity (bottom: $v_h$).}
   \label{fig:obs_astrometry}
\end{figure}

The differential evolution algorithm converged to the solution
$\alpha=149\fdg 25$, $\delta=36\fdg 59$, $D=8.01$ kpc, $\tilde{\mu}_\alpha=-5.56$ mas yr$^{-1}$, $\mu_\delta=-12.40$ mas yr$^{-1}$ and $v_h=-19.15$ km s$^{-1}$, with a value of $\chi^2_{\rm{stream}}= 11.58$.
The orbit that corresponds to these IC is displayed in the observable space in Fig.~\ref{fig:obs_astrometry}~(top: $\phi_2$; middle: $\tilde{\mu}_\alpha$, $\mu_\delta$; bottom: $v_h$) with a dotted (green) line. The solid (black) line shows the corresponding observable data $\eta$, while the shaded (grey) area demarcates the corresponding $1\sigma_\eta$ regions.

\begin{figure}
   \includegraphics[width=\hsize]{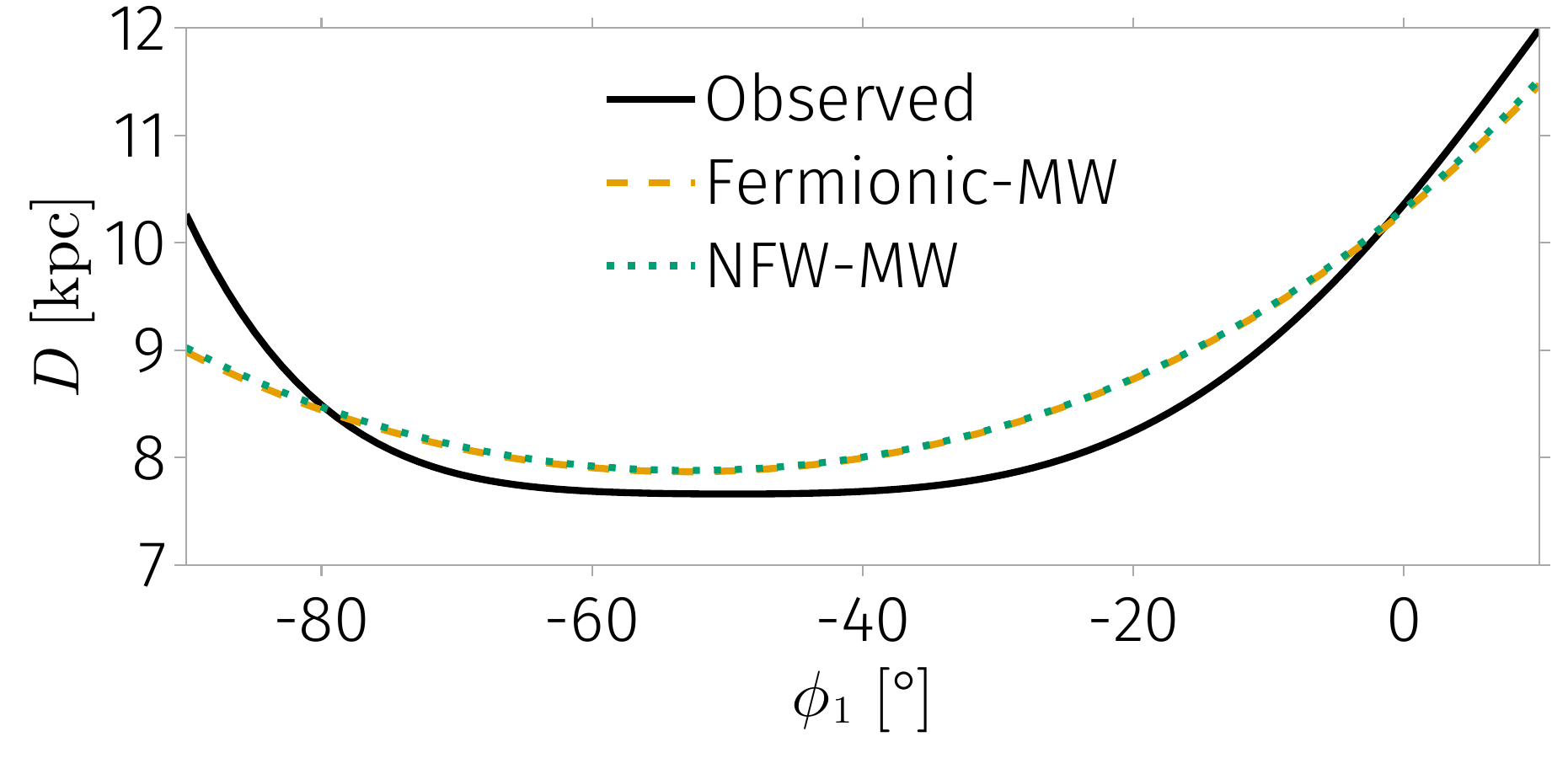}
   \caption{Photometric distance ($D$); not included as an observable in the stream fitting procedure.}
   \label{fig:obs_radial}
\end{figure}
Previous works~\citep{arguelles_novel_2018,2019PDU....24..278A,2023ApJ...945....1K} have shown that the parameters of the fermionic DM model can be split into two sets for the family of fermionic DM profiles with highly degenerate cores (i.e. $\theta_0 \gtrsim 15$ \citealp{2019PDU....24..278A}). On one hand, $m$ and $\beta_0$ control the core of the distribution in the sense that, for given values of $M_{\rm{core}}$ and $m$, it is possible to find a consistent value of $\beta_0$ with a very small influence of the other two parameters $\theta_0$ and $W_0$. This implies a partial degeneracy between $m$ and $\beta_0$.
On the other hand, for the same kind of fermionic \textit{core}-\textit{halo} solutions with positive central degeneracy mentioned above, the main behaviour of the distribution in the halo is determined by $\omega_0$ as explicited in \cite{2019PDU....24..278A} for different galaxy types.

As a starting point for the fitting of the Fermionic-MW parameters (second step), we took the values obtained by \citet{2020A&A...641A..34B}: $(m, \beta_0, \theta_0, W_0)=(56~\rm{keV}~c^{-2}, 1.1977\times10^{-5}, 37.765, 66.3407)$, which allow their MW model (DM+barions) to satisfy the geodesic motions of both S2 and G2 at Sagittarius~A* and the rotation curve from~\citet{sofue_rotation_2013}.
From these values, fixing $m=56~\rm{keV}~c^{-2}$ and taking the already fitted IC of the stream progenitor for the NFW-MW model, we performed a differential evolution minimization of the
$\chi^2_{\rm{full}}$ function in the window
$(\theta_0,\omega_0, \beta_0)\in [35, 40]\times[25, 30]\times[10^{-5}, 1.5\times10^{-5}]$. Using metaparameter values \texttt{maxiter}$~=~$\texttt{popsize}$~=~$300, the algorithm converged to
$(\theta_0, \omega_0, \beta_0)= (36.094, 27.368 , 1.252\times10^{-5})$, giving
$\chi^2_{\rm{stream}}= 16.190$ and  $\chi^2_{\rm{core}}= 7.676\times10^{-10}$.

The third step consisted in polishing the IC of the orbit by using the differential evolution algorithm with fixed Fermionic-MW parameters, with metaparameter values \texttt{maxiter}$~=~$\texttt{popsize}$~=~$400, which gave a result very similar to that of the FMW-MW case:
$\alpha=149\fdg 39$, $\delta=36\fdg 87$, $D=8.02$ kpc,
$\tilde{\mu}_\alpha=-5.55$ mas yr$^{-1}$, $\mu_\delta=-12.33$ mas yr$^{-1}$ and $v_h=-20.84$ km s$^{-1}$, with an improved value of $\chi^2_{\rm{stream}}= 13.59$.

The last step consisted in polishing the fermionic parameters using the second optimization algorithm described in Sec.~\ref{sec:optimization}, i.e. \texttt{NOMAD}. We divided a  macroscopic orthohedron in parameter space, $(\theta_0, \omega_0, \beta_0) \in [35.8, 36.3]\times[27.0, 27.6]\times [1.2\times10^{-5}, 1.3\times10^{-5}]$, in $17^3=4913$ smaller orthoedrons. In each subregion we performed an independent optimization, in a parallel distributed scheme, searching for those parameters that minimize $\chi^2_{\rm{full}}$. Then, we selected the global minimum by comparing the results of each distributed process obtaining the following final fitted parameters of the model: $(\theta_0, \omega_0, \beta_0)= (36.0704, 27.3501, 1.2527\times10^{-5})$, giving $\chi^2_{\rm{full}}= 13.53$.
The corresponding orbit is displayed in the observable space in
Fig.~\ref{fig:obs_astrometry} with a dashed (amber) line. It can be seen that both the Fermionic-MW and NFW-MW models fit the GD-1 stream very well.
We have not performed any statistically rigorous comparison between both models to determine which one is more consistent with the data. In a future paper we will compute the posterior distribution of the fitted parameters, thus being able to give error bounds.

For completeness, in Fig.~\ref{fig:obs_radial} we plot the heliocentric distance ($D$) using the same line types as in Fig.~\ref{fig:obs_astrometry}. The solid (black) line corresponds to the fifth order polynomial fitted in~\cite{Ibata_2020} to the photometric distance there measured:
\begin{align}
 \label{phot_dist}
    \begin{split}
        D &= -4.302\phi_1^5 - 11.54\phi_1^4 - 7.161\phi_1^3 + 5.985\phi_1^2\\
      &\phantom{=} + 8.595\phi_1+10.36.
    \end{split}
\end{align}
It can be seen that both theoretical $D(\phi_1)$ curves agree with each other, but they both differ considerably with respect to the observed one (i.e. the polynomial). Indeed, since the polynomial presents a suspicious constant value for $\phi_1 \in [-70,-30]$ we have plotted the Galactocentric distance versus $\phi_1$ (not shown) and we have found that this latter curve presents an unphysical wobbling behaviour. This fact led us to not include the photometric distance as a fitting observable.

\begin{figure}
   \centering
   \includegraphics[width=\hsize]{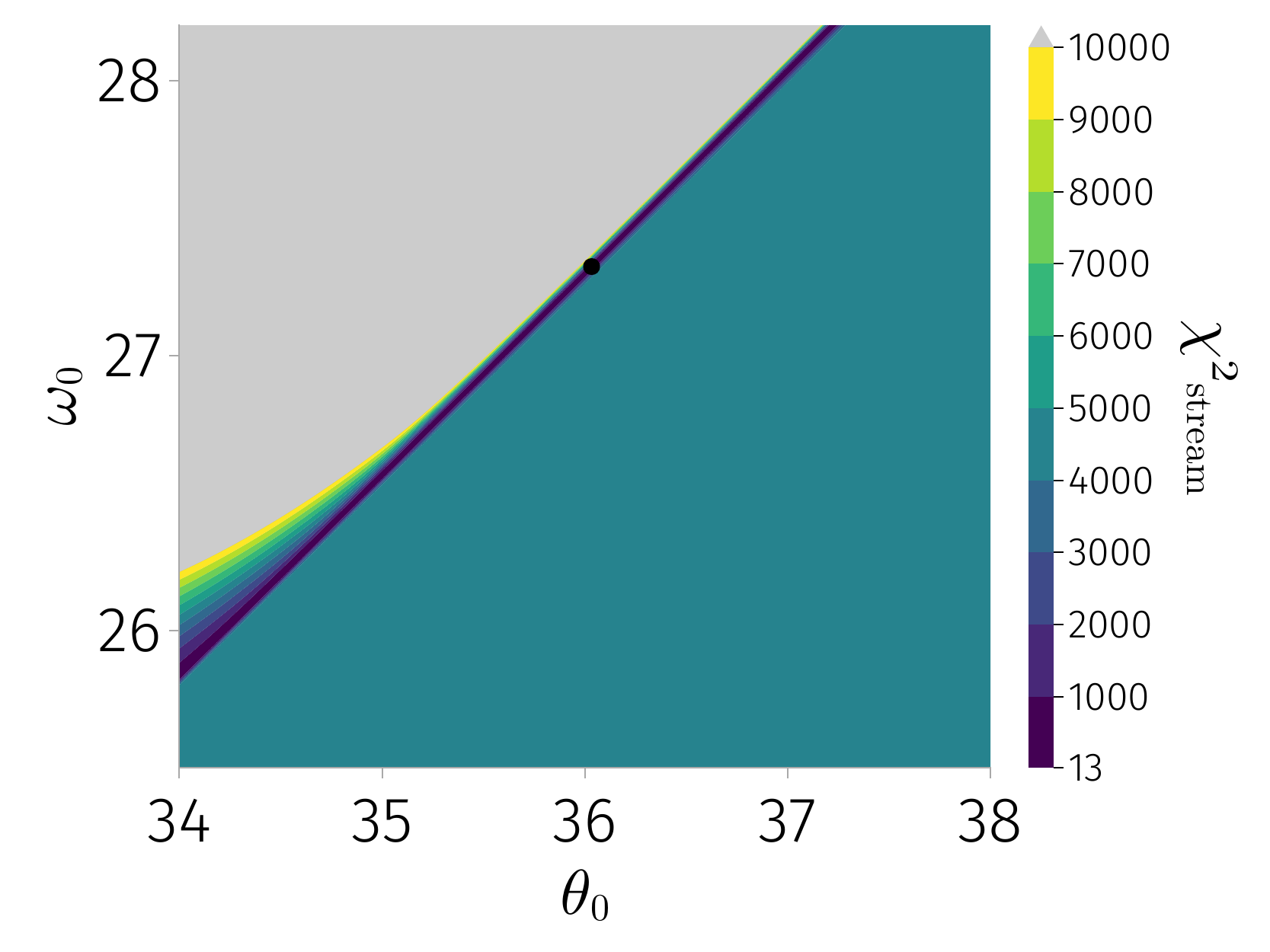}
   \caption{Values of $\chi^2_{\rm{stream}}$ for $(m, \beta_0)=(56~\mathrm{keV}~c^{-2}, 1.254\times10^{-5})$ in the window $(\theta_0, \omega_0)\in[34, 38]\times[25.5,28.2]$. The black point corresponds to our solution and the grey region corresponds to $\chi^2_{\rm{stream}} >10^4$. It can be seen that the minima of the function are located along a thin valley.}
   \label{fig:chi2stream}
\end{figure}

\begin{figure}
   \centering
   \includegraphics[width=\hsize]{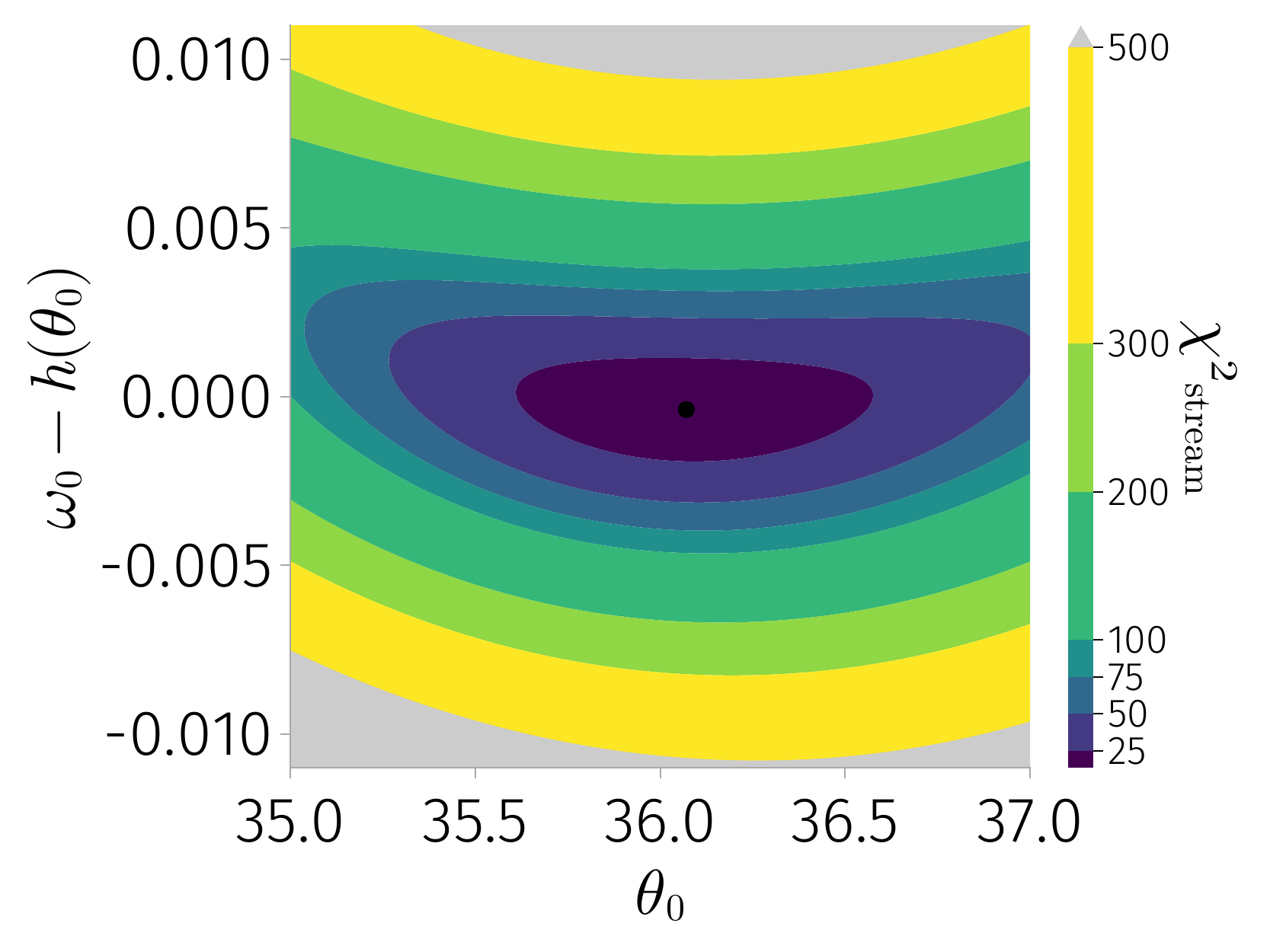}
   \caption{Values of $\chi^2_{\rm{stream}}$ for $(m, \beta_0)=(56~\mathrm{keV}~c^{-2}, 1.254\times10^{-5})$ in
   the window $(\theta_0, \omega_0-h(\theta_0))\in[35, 37]\times[-0.011,0.011]$. The black point corresponds to our solution and the grey region corresponds to $\chi^2_{\rm{stream}} > 500 $. }
   \label{fig:chi2stream_tilted}
\end{figure}

It is instructive to see how the value of $\chi^2_{\rm{stream}}$ is modified when the parameters $\theta_0$ and $\omega_0$ (i.e., those that control the halo) are varied. To this end, we fixed $(m, \beta_0)=(56~\rm{keV}~c^{-2}, 1.2527\times10^{-5})$, and varied $(\theta_0, \omega_0)$ in a grid spanning the rectangle $[34, 38]\times[25, 30]$.
Fig.~\ref{fig:chi2stream} shows a contour plot of this grid coloured by the value of their corresponding $\chi^2_{\rm{stream}}$; the black point corresponds to our fitted solution.
It can be noticed that the minima of the function are located along a thin valley, so the solution of the fitting problem is locally degenerate along a straight line in the $(\theta_0, \omega_0)$ plane.
To see the behaviour of $\chi^2_{\rm{stream}}$ around the solution, we first fitted a line to those points that satisfy $\chi^2_{\rm{stream}}<50$,
obtaining $h(x)= 0.7939+0.7362x$; then subtracted this line to $\omega_0$, and finally remade a plot of $\chi^2_{\rm{stream}}$ in the window $(\theta_0, \omega_0-h(\theta_0))\in[35, 37]\times[-0.011,0.011]$. The result can be seen in Fig.~\ref{fig:chi2stream_tilted}. It is noticeable that indeed there exists a minimum (non-degenerate problem) and that
the variance along $\omega_0-h(\theta_0)$ is two orders of magnitude smaller than the variance along $\theta_0$.

\subsection{Rotation curves, accelerations and virial quantities}
\begin{figure}
   \centering
   \includegraphics[width=\hsize]{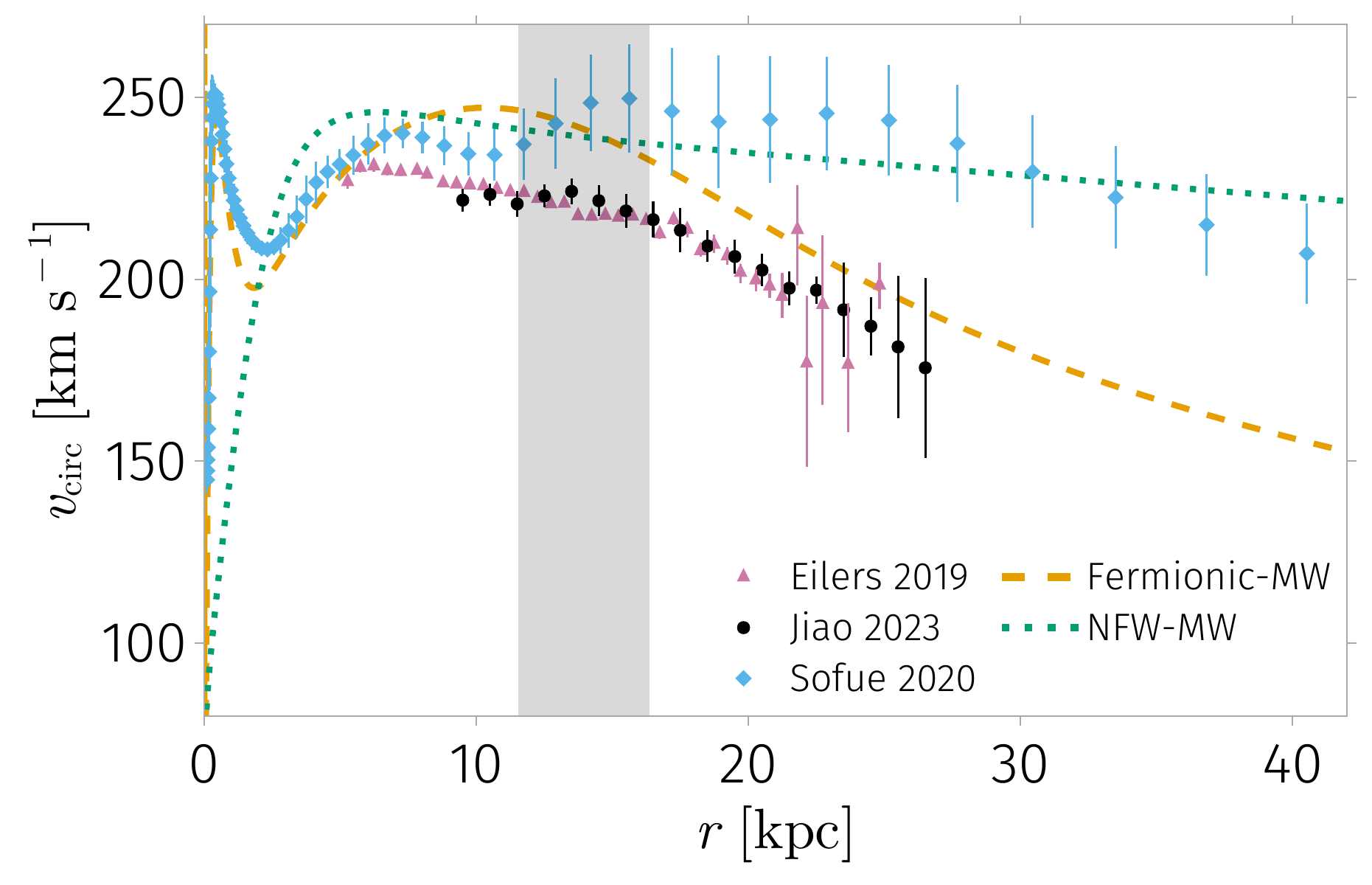}
   \caption{Rotation curves of the Fermionic-MW model (this work, in dashed amber) and NFW-MW model (\cite{2019MNRAS.486.2995M}, in dotted green) which fit the GD-1 stream, are compared \textit{a posteriori} with different observed rotation curves (\citealp{Eilers_2019} with purple triangles,~\citealp{sofue_2020} with light-blue rhombi, and \citealp{Jiao2023} with black circles). Only the Fermionic MW model can account for both, the GD-1 stream data and the sharp drop of the recent GAIA DR3 rotation curve.}
   \label{fig:rotcurve}
\end{figure}

We computed the resulting rotation curves for our two models and compared them with three observed rotation curves (Fig.~\ref{fig:rotcurve}).

The dotted (green) line corresponds to the NFW-MW model and the dashed (amber) one to the Fermionic-MW model.
The observed rotation curves are coded according to: triangles (purple)
for \citet{Eilers_2019}, rhombi (light-blue) for \citet{sofue_2020} and
circles (black) for \citet{Jiao2023}.
In order to build a unified rotation curve, \citet{sofue_2020} computed a running average of many rotation curves resulting from different dynamical tracers according to the Galactocentric distance. In the central parts of the Galaxy, the tracers used were the molecular gas and the infrared stellar motion, while in the outer parts (beyond $r\sim30 \textrm{~kpc}$) the tracers were the radial motions of satellite galaxies and globular clusters. Also rotation curves resulting from galactic disk objects were used.
On the other hand, \citet{Eilers_2019} used a selection of RGB stars as tracers of the disk dynamics.

It is worth noticing  that \citet{sofue_2020} assumes
($R_\odot$,~$v_c(R_\odot))=(8$ kpc, 238 km s$^{-1}$), while \citet{Eilers_2019}
assume $R_\odot=8.122$~kpc and a Galactocentric Sun's velocity $v_\odot$ = $(11.1, 245.8, 7.8)$~km s$^{-1}$,
with which they estimated $v_c(R_\odot)=229\pm0.2$ km s$^{-1}$.

It is interesting to note that both theoretical models, Fermionic-MW and NFW-MW, give $v_\mathrm{c}(R_\odot)\approx 244$~km s$^{-1}$ at the solar radius, which are in excellent agreement with the estimate found
by~\citet{2020arXiv201202169B}, $v_c(R_\odot)=244\pm 8$~km s$^{-1}$ for $R_\odot = 8.275$~kpc, or $v_c(R_\odot)=242\pm 8$~km s$^{-1}$ for our adopted value of $R_\odot$. Although this velocity is larger than the standard value (220-230 km s$^{-1}$), it should be mentioned
that, according to Table 1 in \citet{sofue_2020}, our computed velocity is not an outlier (see also Section 6.2 of~\citealt{Honma_2012}).
Our solution has a an average slope $s=-4.15\pm0.015~\rm{km~s}^{-1}~kpc^{-1}$, fitted for $14.5~\rm{kpc} < r < 26.5~\rm{kpc}$, which is
comparable to the corresponding value of $s=-3.93\pm0.15~\rm{km~s}^{-1}~kpc^{-1}$ measured from the rotation curve of~\citet{Jiao2023}, and in better agreement than the corresponding slope of the NFW-MW model (see also \cref{fig:rotcurve} for comparison).

Using the GD-1 observables, we have computed the present Galactocentric distance projected onto the plane $z=0$, finding that it lies inside the interval 11.5 kpc $\lesssim r \lesssim 16.4$ kpc displayed as a vertical shaded (grey) band in the Galaxy RC of Fig.~\ref{fig:rotcurve}. The GD-1 stream orbit location corresponds to $z\in [2.6, 9.7]$ kpc and thus explores the non-sphericality of the full MW models (due to baryons/NFW axisymmetry).
It is noticeable that both models approximately agree in their circular velocity in the GD-1 region.
The present Galactocentric distance (not projected) corresponds to the interval $13.9~\textrm{kpc} \lesssim r \lesssim 16.6~\rm{kpc}$ (subject to errors in the photometric distance as commented in Section \ref{sec:fitting}). The stream theoretical orbit in the Fermionic-MW has a pericentre of $14.3$~kpc and an apocentre of $24.5$~kpc, and currently it is in its pericentric passage.

We have computed the cylindrical components of the acceleration, $a_r$ and $a_z$,
for both MW models along the stream, obtaining a maximum difference of $|\Delta a_r| \lesssim 0.08 \rm{~km~s}^{-1} \rm{Myr}^{-1}$ and $|\Delta a_r| \lesssim 0.15 \rm{~km~s}^{-1} \rm{Myr}^{-1}$, respectively. This agreement supports  the idea that cold tidal streams are excellent accelerometers~\citep{Ibata_2016,2022ApJ...940...22N,2023ApJ...945L..32C}.

With respect to virial quantities, the \textit{core}-\textit{halo} dark matter solution has a finite virial radius $r_{\rm{DM,vir}}=27.4$ kpc and a virial mass
$M_{\rm{DM,vir}}=1.4\times10^{11} M_\odot$. The total baryon mass of our model is $M_b=0.9\times10^{11}M_\odot$, so the total virial mass amounts to
$M_{\rm{vir}}=2.3\times10^{11} M_\odot$. The value of the MW total mass at 50 kpc reported in Table 3
of \citet{2014MNRAS.445.3788G} is $2.9\times10^{11} M_\odot$ with $(\sigma, 2\sigma)=(0.4,0.9)\times10^{11} M_\odot$, so our solution lies in the $2\sigma$ region.
It should be noted that the mass of the fermionic solution is constant for radii larger than
$r_{\rm{DM,vir}}$, while the mass of the model studied by \citet{2014MNRAS.445.3788G} continues to increase with radius according to their Table 3, though their estimations at large radii have relatively
high error bounds, e.g. $2\sigma=3\times10^{11} M_\odot$ for $M(200 \hbox{ kpc}) =5.6\times10^{11} M_\odot$.

Most recent MW's mass estimations are those obtained by
\citet{Jiao2023} and \citet{Ou2023} from Gaia DR3 data, which report data compatible with even smaller values of the MW virial mass: $1.99^{+0.09}_{-0.06}\times10^{11} M_\odot$ and $2.13^{+0.17}_{-0.12}\times10^{11} M_\odot$ respectively, in agreement with our fermionic model predictions. These MW mass estimates correspond with a sharp Keplerian decline of the MW rotation ending at $\approx 26.5$ kpc (with an enclosed dynamical mass at such radius of $\approx 2\times 10^{11} M_\odot$, \citealp{Jiao2023}), again in remarkable agreement with the virial radius predicted by our Fermionic-MW model of $\approx 27$ kpc.

The fitted fermionic DM model has a density at the solar neighbourhood of
$\rho_{\mathrm{DM},\odot}=1.46\times10^7~M_\odot~\mathrm{kpc}^{-3}=0.55~\mathrm{GeV~cm}^{-3} c^{-2}$ which falls within the $2\sigma$ region of the estimation made by \citealt{Salucci2010} ($0.43\pm 0.21~\mathrm{GeV~cm}^{-3} c^{-2}$)
but is higher than the one obtained by~\citealt{Eilers_2019} ($0.30\pm0.03~\mathrm{GeV~cm}^{-3} c^{-2}$) or by~\citealt{Ou_2024MNRAS} ($0.447\pm0.004~\mathrm{GeV~cm}^{-3} c^{-2}$).

\subsection{An example of the S-cluster fit: the paradigmatic case of the S2 orbit}
In this section we answer the relevant question of how well the fermionic model that fits the stream according to the procedure of \cref{sec:fitting}, can fit the iconic S2 star orbit with focus in Sagittarius~A*. Even if a good fit is expected since the core mass of the DM distribution $M_{\rm{core}}=3.5\times 10^6 M_\odot$ was carefully fitted together with the stream constraint, it is important to notice that the free parameters of the Fermionic-MW model are not precisely the same as the ones obtained in \cite{2020A&A...641A..34B}. That is, while in \cite{2020A&A...641A..34B} the free DM model parameters where obtained to explain the S2 star geodesic together with the MW RC as given in \cite{sofue_rotation_2013}, here the DM halo region was instead fitted in order to reproduce the GD-1 stream, with somewhat different ($\beta_0$,$\theta_0$,$W_0$) values.

\begin{figure}
   \includegraphics[width=1.0\hsize]{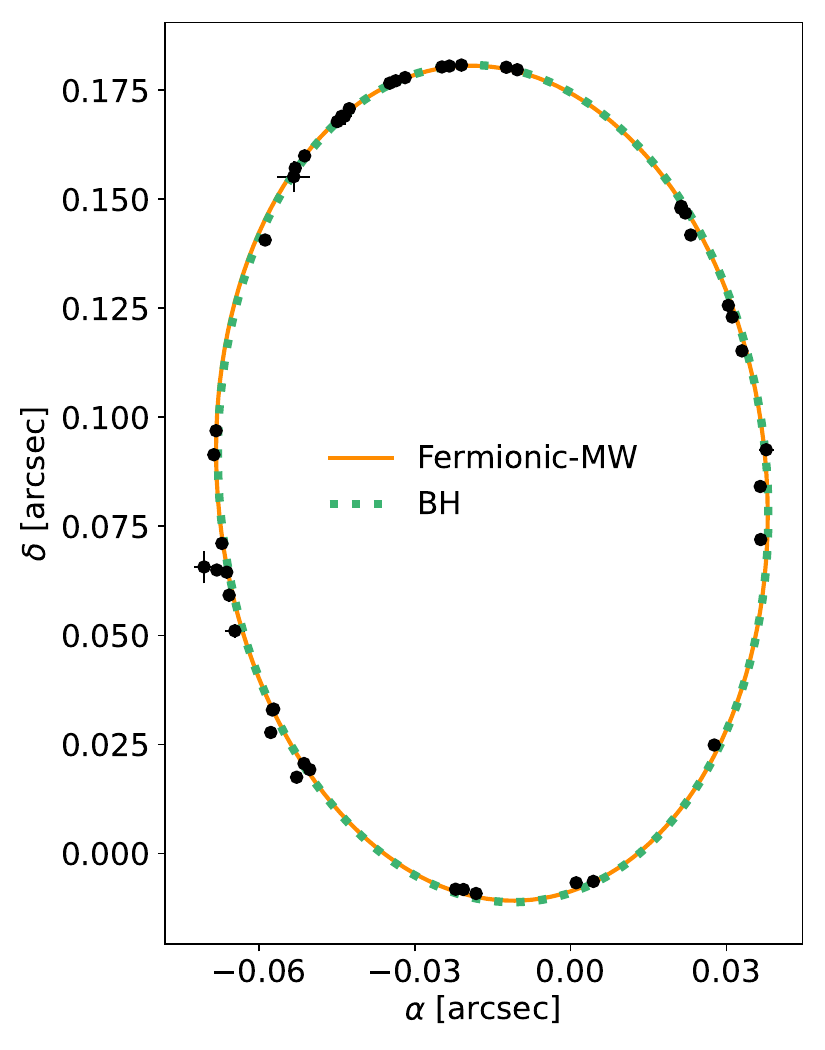}
\caption{Modelled and observed projected orbit in the sky for a Fermionic-MW DM model in a solid (amber) line and a BH of $M_{\rm bh}=4.075\times 10^6 M_\odot$ in a dotted (green) line.}
   \label{fig:S2_fit}
\end{figure}

\begin{figure}
   \includegraphics[width=1.0\hsize]{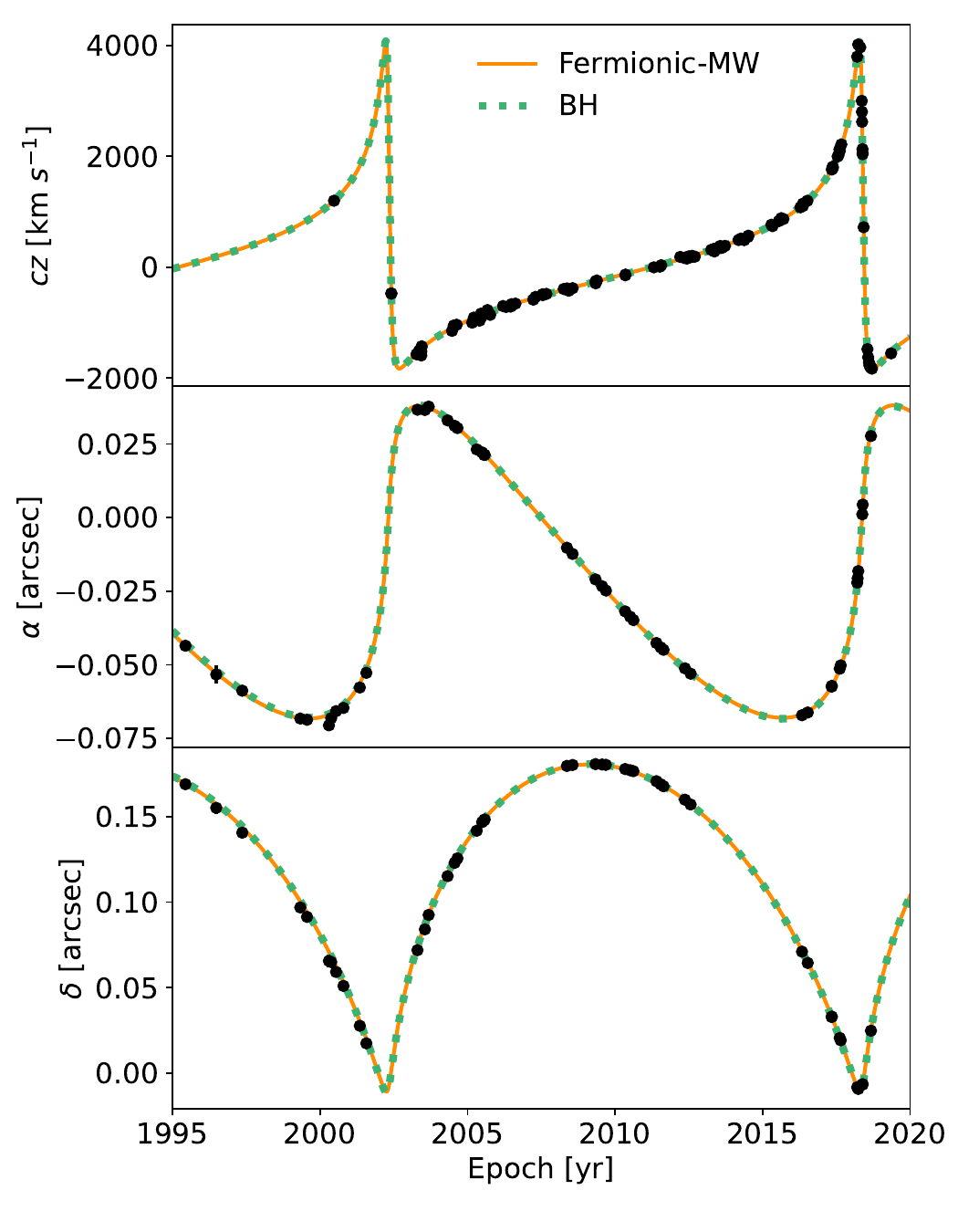}
   \caption{Redshift ($z$), right ascension~($\alpha$) and declination ($\delta$) as a function of time epoch for the same models
   displayed in \cref{fig:S2_fit}.}
   \label{fig:S2_fit_b}
\end{figure}

We have thus performed a least squares fitting procedure following \cite{2020A&A...641A..34B}, for the case of the S2 orbit as an example. As commented above, in this case we use the Fermionic-MW DM model which best fits the GD-1 stream, that is $(\theta_0, \omega_0, \beta_0)= (36.0704, 27.3501, 1.2527\times10^{-5})$, obtaining excellent results. In \cref{fig:S2_fit}  we show the projected S2 orbit in the plane of the sky while in \cref{fig:S2_fit_b} we show the time evolution of the redshift function $z$ (which is directly related to the heliocentric radial velocity according to Equation (C.17a) in~\citealp{2020A&A...641A..34B}), $\alpha$ and $\delta$, for the best fitting values of the \textit{osculating} orbital parameters. These values are given in Table~\ref{tab:S2}, along with the model predicted value of periapsis precession per orbital period, $\Delta \phi$, and the orbital period, $P$.

Our fitting procedure is applied in the gravitational field of two different scenarios: a Fermionic-MW DM model for $m=56~{\rm keV}~c^{-2}$ and $M_{\rm{core}}=3.5\times 10^6 M_\odot$; and a Schwarzschild BH model with central mass of $M_{\rm BH}=4.075\times 10^6 M_\odot$.
The resulting values for the $\chi^2_{\rm{S2}}$ minimization presented in Table~\ref{tab:chi2}, are in perfect agreement with the ones obtained in \cite{2020A&A...641A..34B}\footnote{Following their procedure, we have minimized $\chi^2_{\rm{S2}}$ but we have not computed the posterior distribution of the parameters, lacking the corresponding errors.}.
We have used the latest public accessible data from \cite{2019Sci...365..664D}.

For the exemplified case of the S2 star, a distinction between the models is manifested in the predicted value of the relativistic precession of the S2 periapsis $\Delta \phi$. This interesting relativistic effect in the case of a regular (i.e. non singular) DM core was confronted with the publicly available astrometric data of S2 and carefully compared with the BH case in \cite{2022MNRAS.511L..35A}. There it was shown that larger particle masses (i.e. leading to more compact DM cores as detailed in section below), implies less amount of extended DM mass filling the S2 orbit. Thus, precession growths from retrograde to prograde as it tends to the unique value predicted by the BH model. Indeed in \cite{2022MNRAS.511L..35A} it was shown that already for particle masses of $m = 60~{\rm keV}~c^{-2}$ slightly above the value here considered of $m=56~{\rm keV}~c^{-2}$, the periapsis precession is very close the one predicted by the Schwarzschild BH. On the contrary, for particle masses $m\lesssim 56~{\rm keV}~c^{-2}$ the DM core is too extended in radius producing large values of retrograde S2 periapsis precesion and poorer orbit fits \citep{2020A&A...641A..34B,2022MNRAS.511L..35A}.

\begin{table}[t]
\caption{Best-fitting \textit{osculating} orbital parameters of the orbit of the S2 star for two different models: a fermionic DM model with $m=56~{\rm keV}~c^{-2}$, core mass $M_{\rm{core}}=3.5\times 10^6 M_\odot$, and a BH of mass  $M_{\rm bh}=4.075\times 10^6 M_\odot$.}
\centering
\begin{tabular}{lcc}
\hline
Parameter & Fermionic-MW & BH \\
\hline
$a$ [as] & 0.12507 & 0.12530 \\
$e$ & 0.8868 & 0.8861 \\
$\omega$ [$^{\circ}$] & 66.935 & 66.505 \\
$i$ [$^{\circ}$] & 134.396 & 134.440 \\
$\Omega$ [$^{\circ}$] & 228.195 & 228.046 \\
$P$ [yr] & 16.051 & 16.049 \\
$\Delta \phi$ [arcmin rev$^{-1}$] & -6.04 & 11.95 \\
\hline
\end{tabular}
\label{tab:S2}
\end{table}

\subsection{Varying the fermion mass to reach more compact cores}
\begin{figure}
   \centering
   \includegraphics[width=\hsize]{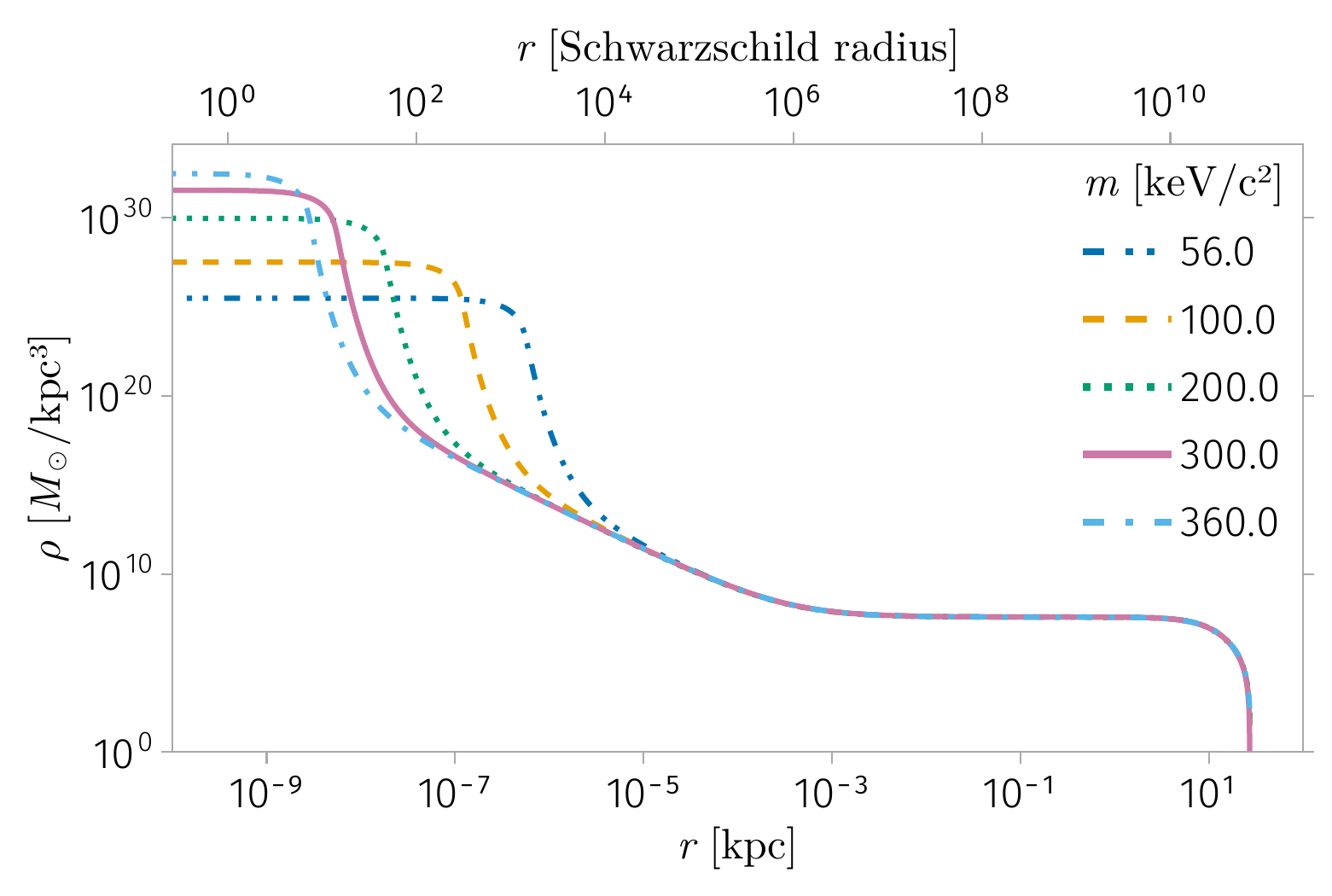}
   \caption{Fermionic DM density profiles with different core-compacities (i.e. different $m$) fitting both, the GD-1 stream and the DM core mass which agree with the S2 star data orbiting Sagittarius A*. The Schwarzschild radius is computed assuming a black hole mass $M_{\rm bh}=\SI{4.075E6}{\Msun}$.}
   \label{fig:going_compact}
\end{figure}

As already mentioned, we have found a fermionic solution that is in agreement with both GD-1 data and
the geodesic motion of the best studied S-cluster star around Sagittarius~A*, the S2 star. But it is also true that this S-stars constraint will be also satisfied by any fermionic DM profile
with a core mass $m_\mathrm{c}\approx M_{\rm{core}}$ and more compact cores than the solution corresponding for $m=56$ keV. It is therefore interesting to find out how much compactness can be reached
while keeping both the S2 star and \hbox{GD-1} constraints, in the light of the new observations
of the Event Horizon Telescope, \cite{EHT_image}, where a shadow angular diameter of $48.7\pm7.0~\mu$as
has been measured. This diameter corresponds to a shadow radius of $\sim 2.46$ Schwarzschild radii assuming
a black hole mass of $M_{\rm{bh}}=4.075\times10^6~M_\odot$.
In order to extend the fermionic solutions to other values of the fermion mass ($m$), we used the second optimization algorithm described in Sec.~\ref{sec:optimization} for $m=100$, 200, 300 and 360 keV $c^{-2}$. For each fermion mass we divided a given macroscopic orthohedron\footnote{The lower and upper bounds of the orthohedrons were given by
lower = $(36, 27, 1.2\times10^{-5})$, $(37, 28, 5\times10^{-5})$, $(38, 29, 3.5\times10^{-4})$, $(40, 29, 1.3\times10^{-3})$ and
upper = $(40, 31, 10^{-4})$, $(41, 32, 10^{-3})$, $(42, 32, 3\times10^{-3})$, $(44, 32, 4\times10^{-3})$, respectively for $m=100$, 200, 300, 360 keV $c^{-2}$ in $(\theta_0, \omega_0, \beta_0)$ space.}
in parameter space in $20^3=8000$ smaller orthoedrons. In each subregion we performed an independent optimization in a parallel distributed scheme, searching for those parameters that minimize $\chi^2_{\rm{full}}$ with the \texttt{NOMAD} algorithm. Afterwards, we selected the global minimum by comparing the results of each distributed process.
The result is that for all the fermion masses it is possible to find values of the other parameters in such a way that both the GD-1 stream and the core mass constraints  are respected with the same precision as in the initial ($m=56~\rm{keV}/c^2$) case. In fact, as shown in Fig.~\ref{fig:going_compact}, all the solutions have the same density profile in the halo region, while their difference is limited to the compactness of the core.
In Table~\ref{tab:chi2} we show the values of $\chi^2_{\rm{stream}}$
for all the fermion masses studied. We also give the value of $\chi^2_{\rm{S2}}$ for
the Fermionic-MW and the BH models. It is seen that all the fermionic models analyzed here are statistically indistinguishable\footnote{These fits were done for the same fixed core mass as in~\cite{2020A&A...641A..34B, 2021MNRAS.505L..64B}, but it can be shown that the value of $\chi^2_{\rm{S2}}$ corresponding to $m=360~\rm{keV}$ can be made as small as in the other cases, by increasing the core mass a few percent.}, and more data (e.g. central shadow feature, or closer/fainter S-stars to SgrA* than S-2) are needed in order to further constrain the particle mass range. Both projects are currently under development within our group.
The values of the core radii of these solutions are approximately 1097, 232, 35, 10, 5 Schwarzschild radii for $m=56$, 100, 200, 300, 360 keV $c^{-2}$, respectively. The last analysed value of $m=360$ keV $c^{-2}$ corresponds to a DM core very close to the last stable solution according to the stability criterium of \cite{2021MNRAS.502.4227A}, leading to a gravitational core-collapse into a BH of about $4\times 10^6 M_\odot$.

\begin{table}[t]
\caption{$\chi^2_{\rm{stream}}$ and $\chi^2_{\rm{S2}}$ values corresponding to the best-fit to the GD-1 stream and the S2 orbit, respectively.}
\centering
\small{
\begin{tabular}{lcccccc}
\hline
  $mc^2/$keV & 56 & 100 & 200 & 300 & 360 & BH \\
\hline \hline
$\chi^2_{\rm{stream}}$ & 13.528 & 13.530  & 13.575 & 13.862  & 13.836  \\
\hline
$\chi^2_{\rm{S2}}$ & 3.185 & 3.405 & 3.349 & 4.371 & 12.561 & 3.383 \\ \hline
\end{tabular}
}
\label{tab:chi2}
\end{table}
\section{Conclusions}
\label{sec:conclusions}
We have fitted both, the GD-1 stream located at about $14-15$ kpc from the Galaxy center together with the S2 star orbit located at miliparsec scales, in a MW potential consisting of a fermionic \textit{core}-\textit{halo} DM distribution \citep{arguelles_novel_2018,2020A&A...641A..34B,2021MNRAS.505L..64B,2022MNRAS.511L..35A}, plus a fixed baryonic distribution \citep{2017A&A...598A..66P}. Remarkably, the resulting MW total mass and virial radius of the Galaxy predicted by the fermionic DM model is in excellent agreement with both, the virial mass of $\approx 2\times 10^{11} M_\odot$ and the Galactocentric radial range of $\approx 20-26$ kpc in which the MW rotation curve sharply drops as recently meassured by Gaia DR3. This relatively low mass MW is also consistent with the independent estimate given by~\cite{2014MNRAS.445.3788G}, though it is considerably lower than typical values given in previous literature, e.g.~\cite{2010MNRAS.406..264W} and references therein.

We obtained the free parameters of the fermionic model by fixing the fermion mass and fitting simultaneously two astrophysical constraints: the stream observables and a
DM core mass of $3.5\times10^6M_\odot$, the latter taken from previous fits of the S-stars cluster at
the centre of the Galaxy with no central BH~\citep{2020A&A...641A..34B,2021MNRAS.505L..64B}. We could thus reproduce with high accuracy, and for fermion masses ranging from 56 to 360~keV~$c^{-2}$, the polynomials fitted by~\citet{Ibata_2020} that correspond to observed sky position, proper motion, photometric distance and radial velocity of the stream.

In order to compare with other GD-1 fits in the literature, we have also fitted the stream progenitor initial conditions with the  axisymmetric generalization of the NFW distribution from~\citet{2019MNRAS.486.2995M}, obtaining agreement in the GD-1 phase-space track of
both the Fermionic-MW and the NFW-MW models. Additionally, we have obtained agreement between both models in the rotation curves at projected ($z=0$) Galactocentric distances, $r$, corresponding to the stream observables, i.e. $11.5\leq r \leq 16.4$ kpc. The average slope of the rotation curve between $14.5$ and $26.5$ kpc obtained from the Fermionic-MW model was $s=-4.18 \pm 0.02~\rm{km\, s}^{-1} kpc^{-1}$, in much better agreement with the recent observations of \cite{Jiao2023} than the NFW-MW model. Regarding the acceleration field, we have shown that both MW models perfectly agree in their acceleration vectors as a function of the position along the stream.

We have found a circular velocity at the Sun's position of $v_c(R_\odot)= 244~\rm{km~s}^{-1}$, in line with the value independently obtained by \citet{2019MNRAS.486.2995M}.

The fermionic DM solution has a finite radius of $r_{\rm{DM,vir}}=27.4$~kpc and a virial mass of
$M_{\rm{DM,vir}}=1.4\times10^{11}~M_\odot$, implying a total (DM plus baryons) virial mass of the galaxy of $M_{\rm{vir}}= 2.4\times10^{11}~M_\odot$, which is at $2\sigma$ from the value reported in Table 3 of
\citet{2014MNRAS.445.3788G} for a radius of 50~kpc. The value obtained for the DM density at the solar neighbourhood is $\rho_{DM,\odot}=1.46\times10^7M_\odot~\mathrm{kpc}^{-3}=0.55~\mathrm{GeV}~\mathrm{cm}^{-3} c^{-2}$, which falls inside the $2\sigma$ region of a previous estimate by \citet{Salucci2010}.

Finally, we have shown that it is possible to find a one-dimensional family of solutions parameterized by the fermion mass, having the same halo that fits the GD-1 stream but differing in the compactness of the central core, and always reproducing the S2 star orbit (see \cref{fig:S2_fit,fig:S2_fit_b} for the case of $m=56$ keV). For the limiting case studied here ($m=360$~keV) we obtained a core radius of $r_\mathrm{c}\approx5$~Schwarzschild radii.
A precise relativistic ray tracing study about simulated ring-like images of the central
cores of fermionic distributions is on progress, trying to put strict constraints on the minimum
compactness needed to be in agreement with the EHT observations.

In summary, the findings of this work not only support the idea that cold tidal streams are excellent probes of the acceleration field of the Milky Way, but show that the (spherical) fermionic model is capable to fit independent set of observables covering three totally different Galaxy distance scales: $\sim 10^{-6}$ kpc (S-cluster), $\sim 14$ kpc (GD-1) and $\sim 30$ kpc (GAIA DR3 RC mass estimates).

\begin{acknowledgements}
    MFM would like to dedicate this work to the memory of a great friend since childhood, Gustavo Fabián Larrion (1980-2021), Q.E.P.D.
    We thank the anonymous referee for helping to improve the paper.
    MFM thanks Khyati Malhan for his comments about the NFW-MW model.
    MFM thanks Leandro Martínez, Ian Weaver, Joaquín Pelle and the \textsc{Julia} community for their
    great support with the language and workflow. We thank Jorge A. Rueda for his comments
    about the manuscript.
    We thank Juan Ignacio Rodriguez for his great support with hardware and software issues regarding the IALP server and personal computers. We also thank Federico Bareilles and the informatic support team of the FCAGLP for their constant help.
    This work used computational resources from CCAD – Universidad
    Nacional de Córdoba (\href{https://ccad.unc.edu.ar/}{https://ccad.unc.edu.ar/}),
    which are part of SNCAD – MinCyT, República Argentina.
    This work also used computational resources from the HPC center DIRAC,
    funded by Instituto de Física de Buenos Aires (UBA-CONICET) and part of SNCAD-MinCyT initiative, Argentina.
    DDC and MFM acknowledge support from CONICET (PIP2169) and from the Universidad Nacional de La Plata (G168). C.R.A. acknowledges support from CONICET, the ANPCyT (grant PICT-2018-03743), and ICRANet. VC thanks financial support from CONICET, Argentina.
    The figures presented in this work where made with the
    {\it AlgebraOfGraphics.jl}~(\href{https://aog.makie.org/dev/}{https://aog.makie.org/dev/}),
    {\it Makie.jl}~\citep{DanischKrumbiegel2021} and {\it Matplotlib}~\citep{Hunter:2007}
    packages. Some of our optimization results were initially guided by the use of the \texttt{LN\_NELDERMEAD}~\citep{NELDERMEAD,10.1093/comjnl/8.1.42,shere} algorithm, from the {\it NLopt.jl}~\citep{NLopt} package.
    In order to run {\sc Julia} in a parallel SLURM environment we made use of the {\it Distributed.jl} and \href{https://github.com/kleinhenz/SlurmClusterManager.jl}{\it SlurmClusterManager.jl} packages.
\end{acknowledgements}

%
%
\bibliographystyle{aa} 
\bibliography{refs} 

\begin{thebibliography}{85}
\expandafter\ifx\csname natexlab\endcsname\relax\def\natexlab#1{#1}\fi

\bibitem[{{Amorisco}(2015)}]{2015MNRAS.450..575A}
{Amorisco}, N.~C. 2015, \mnras, 450, 575

\bibitem[{{Arg{\"u}elles} {et~al.}(2021){Arg{\"u}elles}, {D{\'\i}az}, {Krut},
  \& {Yunis}}]{2021MNRAS.502.4227A}
{Arg{\"u}elles}, C.~R., {D{\'\i}az}, M.~I., {Krut}, A., \& {Yunis}, R. 2021,
  \mnras, 502, 4227

\bibitem[{{Arg{\"u}elles} {et~al.}(2019){Arg{\"u}elles}, {Krut}, {Rueda}, \&
  {Ruffini}}]{2019PDU....24..278A}
{Arg{\"u}elles}, C.~R., {Krut}, A., {Rueda}, J.~A., \& {Ruffini}, R. 2019,
  Physics of the Dark Universe, 24, 100278

\bibitem[{{Arg{\"u}elles} {et~al.}(2022){Arg{\"u}elles}, {Mestre},
  {Becerra-Vergara}, {Crespi}, {Krut}, {Rueda}, \&
  {Ruffini}}]{2022MNRAS.511L..35A}
{Arg{\"u}elles}, C.~R., {Mestre}, M.~F., {Becerra-Vergara}, E.~A., {et~al.}
  2022, \mnras, 511, L35

\bibitem[{Argüelles {et~al.}(2018)Argüelles, Krut, Rueda, \&
  Ruffini}]{arguelles_novel_2018}
Argüelles, C.~R., Krut, A., Rueda, J.~A., \& Ruffini, R. 2018, Physics of the
  Dark Universe, 21

\bibitem[{{Astropy Collaboration} {et~al.}(2022){Astropy Collaboration},
  {Price-Whelan}, {Lim}, {Earl}, {Starkman}, {Bradley}, {Shupe}, {Patil},
  {Corrales}, {Brasseur}, {N{"o}the}, {Donath}, {Tollerud}, {Morris},
  {Ginsburg}, {Vaher}, {Weaver}, {Tocknell}, {Jamieson}, {van Kerkwijk},
  {Robitaille}, {Merry}, {Bachetti}, {G{"u}nther}, {Aldcroft},
  {Alvarado-Montes}, {Archibald}, {B{'o}di}, {Bapat}, {Barentsen}, {Baz{'a}n},
  {Biswas}, {Boquien}, {Burke}, {Cara}, {Cara}, {Conroy}, {Conseil}, {Craig},
  {Cross}, {Cruz}, {D'Eugenio}, {Dencheva}, {Devillepoix}, {Dietrich},
  {Eigenbrot}, {Erben}, {Ferreira}, {Foreman-Mackey}, {Fox}, {Freij}, {Garg},
  {Geda}, {Glattly}, {Gondhalekar}, {Gordon}, {Grant}, {Greenfield}, {Groener},
  {Guest}, {Gurovich}, {Handberg}, {Hart}, {Hatfield-Dodds}, {Homeier},
  {Hosseinzadeh}, {Jenness}, {Jones}, {Joseph}, {Kalmbach}, {Karamehmetoglu},
  {Ka{l}uszy{'n}ski}, {Kelley}, {Kern}, {Kerzendorf}, {Koch}, {Kulumani},
  {Lee}, {Ly}, {Ma}, {MacBride}, {Maljaars}, {Muna}, {Murphy}, {Norman},
  {O'Steen}, {Oman}, {Pacifici}, {Pascual}, {Pascual-Granado}, {Patil},
  {Perren}, {Pickering}, {Rastogi}, {Roulston}, {Ryan}, {Rykoff}, {Sabater},
  {Sakurikar}, {Salgado}, {Sanghi}, {Saunders}, {Savchenko}, {Schwardt},
  {Seifert-Eckert}, {Shih}, {Jain}, {Shukla}, {Sick}, {Simpson},
  {Singanamalla}, {Singer}, {Singhal}, {Sinha}, {Sip{H{o}}cz}, {Spitler},
  {Stansby}, {Streicher}, {{{S}}umak}, {Swinbank}, {Taranu}, {Tewary},
  {Tremblay}, {Val-Borro}, {Van Kooten}, {Vasovi{'c}}, {Verma}, {de Miranda
  Cardoso}, {Williams}, {Wilson}, {Winkel}, {Wood-Vasey}, {Xue}, {Yoachim},
  {Zhang}, {Zonca}, \& {Astropy Project Contributors}}]{astropy:2022}
{Astropy Collaboration}, {Price-Whelan}, A.~M., {Lim}, P.~L., {et~al.} 2022,
  apj, 935, 167

\bibitem[{{Astropy Collaboration} {et~al.}(2018){Astropy Collaboration},
  {Price-Whelan}, {Sip{\H{o}}cz}, {G{\"u}nther}, {Lim}, {Crawford}, {Conseil},
  {Shupe}, {Craig}, {Dencheva}, {Ginsburg}, {Vand erPlas}, {Bradley},
  {P{\'e}rez-Su{\'a}rez}, {de Val-Borro}, {Aldcroft}, {Cruz}, {Robitaille},
  {Tollerud}, {Ardelean}, {Babej}, {Bach}, {Bachetti}, {Bakanov}, {Bamford},
  {Barentsen}, {Barmby}, {Baumbach}, {Berry}, {Biscani}, {Boquien}, {Bostroem},
  {Bouma}, {Brammer}, {Bray}, {Breytenbach}, {Buddelmeijer}, {Burke},
  {Calderone}, {Cano Rodr{\'\i}guez}, {Cara}, {Cardoso}, {Cheedella}, {Copin},
  {Corrales}, {Crichton}, {D'Avella}, {Deil}, {Depagne}, {Dietrich}, {Donath},
  {Droettboom}, {Earl}, {Erben}, {Fabbro}, {Ferreira}, {Finethy}, {Fox},
  {Garrison}, {Gibbons}, {Goldstein}, {Gommers}, {Greco}, {Greenfield},
  {Groener}, {Grollier}, {Hagen}, {Hirst}, {Homeier}, {Horton}, {Hosseinzadeh},
  {Hu}, {Hunkeler}, {Ivezi{\'c}}, {Jain}, {Jenness}, {Kanarek}, {Kendrew},
  {Kern}, {Kerzendorf}, {Khvalko}, {King}, {Kirkby}, {Kulkarni}, {Kumar},
  {Lee}, {Lenz}, {Littlefair}, {Ma}, {Macleod}, {Mastropietro}, {McCully},
  {Montagnac}, {Morris}, {Mueller}, {Mumford}, {Muna}, {Murphy}, {Nelson},
  {Nguyen}, {Ninan}, {N{\"o}the}, {Ogaz}, {Oh}, {Parejko}, {Parley}, {Pascual},
  {Patil}, {Patil}, {Plunkett}, {Prochaska}, {Rastogi}, {Reddy Janga},
  {Sabater}, {Sakurikar}, {Seifert}, {Sherbert}, {Sherwood-Taylor}, {Shih},
  {Sick}, {Silbiger}, {Singanamalla}, {Singer}, {Sladen}, {Sooley},
  {Sornarajah}, {Streicher}, {Teuben}, {Thomas}, {Tremblay}, {Turner},
  {Terr{\'o}n}, {van Kerkwijk}, {de la Vega}, {Watkins}, {Weaver}, {Whitmore},
  {Woillez}, {Zabalza}, \& {Astropy Contributors}}]{astropy:2018}
{Astropy Collaboration}, {Price-Whelan}, A.~M., {Sip{\H{o}}cz}, B.~M., {et~al.}
  2018, \aj, 156, 123

\bibitem[{{Astropy Collaboration} {et~al.}(2013){Astropy Collaboration},
  {Robitaille}, {Tollerud}, {Greenfield}, {Droettboom}, {Bray}, {Aldcroft},
  {Davis}, {Ginsburg}, {Price-Whelan}, {Kerzendorf}, {Conley}, {Crighton},
  {Barbary}, {Muna}, {Ferguson}, {Grollier}, {Parikh}, {Nair}, {Unther},
  {Deil}, {Woillez}, {Conseil}, {Kramer}, {Turner}, {Singer}, {Fox}, {Weaver},
  {Zabalza}, {Edwards}, {Azalee Bostroem}, {Burke}, {Casey}, {Crawford},
  {Dencheva}, {Ely}, {Jenness}, {Labrie}, {Lim}, {Pierfederici}, {Pontzen},
  {Ptak}, {Refsdal}, {Servillat}, \& {Streicher}}]{astropy:2013}
{Astropy Collaboration}, {Robitaille}, T.~P., {Tollerud}, E.~J., {et~al.} 2013,
  \aap, 558, A33

\bibitem[{Audet \& Dennis(2006)}]{MADS_2006}
Audet, C. \& Dennis, J.~E. 2006, SIAM Journal on Optimization, 17, 188

\bibitem[{Audet {et~al.}(2021)Audet, Digabel, Montplaisir, \&
  Tribes}]{audet2021nomad}
Audet, C., Digabel, S.~L., Montplaisir, V.~R., \& Tribes, C. 2021, NOMAD
  version 4: Nonlinear optimization with the MADS algorithm

\bibitem[{{Becerra-Vergara} {et~al.}(2020){Becerra-Vergara}, {Arg{\"u}elles},
  {Krut}, {Rueda}, \& {Ruffini}}]{2020A&A...641A..34B}
{Becerra-Vergara}, E.~A., {Arg{\"u}elles}, C.~R., {Krut}, A., {Rueda}, J.~A.,
  \& {Ruffini}, R. 2020, \aap, 641, A34

\bibitem[{{Becerra-Vergara} {et~al.}(2021){Becerra-Vergara}, {Arg{\"u}elles},
  {Krut}, {Rueda}, \& {Ruffini}}]{2021MNRAS.505L..64B}
{Becerra-Vergara}, E.~A., {Arg{\"u}elles}, C.~R., {Krut}, A., {Rueda}, J.~A.,
  \& {Ruffini}, R. 2021, \mnras, 505, L64

\bibitem[{Bezanson {et~al.}(2017)Bezanson, Edelman, Karpinski, \&
  Shah}]{bezanson2017julia}
Bezanson, J., Edelman, A., Karpinski, S., \& Shah, V.~B. 2017, SIAM review, 59,
  65

\bibitem[{{Bovy}(2015)}]{2015ApJS..216...29B}
{Bovy}, J. 2015, \apjs, 216, 29

\bibitem[{{Bovy}(2020)}]{2020arXiv201202169B}
{Bovy}, J. 2020, arXiv e-prints, arXiv:2012.02169

\bibitem[{Box(1965)}]{10.1093/comjnl/8.1.42}
Box, M.~J. 1965, The Computer Journal, 8, 42

\bibitem[{{Carlberg}(2018)}]{2018ApJ...861...69C}
{Carlberg}, R.~G. 2018, \apj, 861, 69

\bibitem[{{Chavanis}(2004)}]{2004PhyA..332...89C}
{Chavanis}, P.-H. 2004, Physica A Statistical Mechanics and its Applications,
  332, 89

\bibitem[{{Chavanis}(2020)}]{2020EPJP..135..290C}
{Chavanis}, P.-H. 2020, European Physical Journal Plus, 135, 290

\bibitem[{{Chavanis}(2022{\natexlab{a}})}]{2022PhyA..60628089C}
{Chavanis}, P.-H. 2022{\natexlab{a}}, Physica A Statistical Mechanics and its
  Applications, 606, 128089

\bibitem[{{Chavanis}(2022{\natexlab{b}})}]{2022PhRvD.106d3538C}
{Chavanis}, P.-H. 2022{\natexlab{b}}, \prd, 106, 043538

\bibitem[{{Chavanis} {et~al.}(2015){Chavanis}, {Lemou}, \&
  {M{\'e}hats}}]{2015PhRvD..92l3527C}
{Chavanis}, P.-H., {Lemou}, M., \& {M{\'e}hats}, F. 2015, \prd, 92, 123527

\bibitem[{{Collaboration, T.E.H.T}(2022)}]{EHT_image}
{Collaboration, T.E.H.T}. 2022, The Astrophysical Journal Letters, 930

\bibitem[{{Craig} {et~al.}(2023){Craig}, {Chakrabarti}, {Sanderson}, \&
  {Nikakhtar}}]{2023ApJ...945L..32C}
{Craig}, P., {Chakrabarti}, S., {Sanderson}, R.~E., \& {Nikakhtar}, F. 2023,
  \apjl, 945, L32

\bibitem[{{Cunningham} {et~al.}(2023){Cunningham}, {Hunt}, {Price-Whelan},
  {Johnston}, {Ness}, {Lu}, {Escala}, \& {Stelea}}]{2023arXiv230708730C}
{Cunningham}, E.~C., {Hunt}, J. A.~S., {Price-Whelan}, A.~M., {et~al.} 2023,
  arXiv e-prints, arXiv:2307.08730

\bibitem[{Danisch \& Krumbiegel(2021)}]{DanischKrumbiegel2021}
Danisch, S. \& Krumbiegel, J. 2021, Journal of Open Source Software, 6, 3349

\bibitem[{de~Boer {et~al.}(2018)de~Boer, Belokurov, Koposov, Ferrarese, Erkal,
  Côté, \& Navarro}]{10.1093/mnras/sty677}
de~Boer, T. J.~L., Belokurov, V., Koposov, S.~E., {et~al.} 2018, Monthly
  Notices of the Royal Astronomical Society, 477, 1893

\bibitem[{{Do} {et~al.}(2019){Do}, {Hees}, {Ghez}, {Martinez}, {Chu}, {Jia},
  {Sakai}, {Lu}, {Gautam}, {O'Neil}, {Becklin}, {Morris}, {Matthews},
  {Nishiyama}, {Campbell}, {Chappell}, {Chen}, {Ciurlo}, {Dehghanfar},
  {Gallego-Cano}, {Kerzendorf}, {Lyke}, {Naoz}, {Saida}, {Sch{\"o}del},
  {Takahashi}, {Takamori}, {Witzel}, \& {Wizinowich}}]{2019Sci...365..664D}
{Do}, T., {Hees}, A., {Ghez}, A., {et~al.} 2019, Science, 365, 664

\bibitem[{Eilers {et~al.}(2019)Eilers, Hogg, Rix, \& Ness}]{Eilers_2019}
Eilers, A.-C., Hogg, D.~W., Rix, H.-W., \& Ness, M.~K. 2019, The Astrophysical
  Journal, 871, 120

\bibitem[{{Gialluca} {et~al.}(2021){Gialluca}, {Naidu}, \&
  {Bonaca}}]{2021ApJ...911L..32G}
{Gialluca}, M.~T., {Naidu}, R.~P., \& {Bonaca}, A. 2021, \apjl, 911, L32

\bibitem[{{Gibbons} {et~al.}(2014){Gibbons}, {Belokurov}, \&
  {Evans}}]{2014MNRAS.445.3788G}
{Gibbons}, S.~L.~J., {Belokurov}, V., \& {Evans}, N.~W. 2014, \mnras, 445, 3788

\bibitem[{{GRAVITY Collaboration} {et~al.}(2018){GRAVITY Collaboration},
  {Abuter}, {Amorim}, {Anugu}, {Baub{\"o}ck}, {Benisty}, {Berger}, {Blind},
  {Bonnet}, {Brandner}, {Buron}, {Collin}, {Chapron}, {Cl{\'e}net}, {Coud{\'e}
  Du Foresto}, {de Zeeuw}, {Deen}, {Delplancke-Str{\"o}bele}, {Dembet},
  {Dexter}, {Duvert}, {Eckart}, {Eisenhauer}, {Finger}, {F{\"o}rster
  Schreiber}, {F{\'e}dou}, {Garcia}, {Garcia Lopez}, {Gao}, {Gendron},
  {Genzel}, {Gillessen}, {Gordo}, {Habibi}, {Haubois}, {Haug}, {Hau{\ss}mann},
  {Henning}, {Hippler}, {Horrobin}, {Hubert}, {Hubin}, {Jimenez Rosales},
  {Jochum}, {Jocou}, {Kaufer}, {Kellner}, {Kendrew}, {Kervella}, {Kok},
  {Kulas}, {Lacour}, {Lapeyr{\`e}re}, {Lazareff}, {Le Bouquin}, {L{\'e}na},
  {Lippa}, {Lenzen}, {M{\'e}rand}, {M{\"u}ler}, {Neumann}, {Ott}, {Palanca},
  {Paumard}, {Pasquini}, {Perraut}, {Perrin}, {Pfuhl}, {Plewa}, {Rabien},
  {Ram{\'\i}rez}, {Ramos}, {Rau}, {Rodr{\'\i}guez-Coira}, {Rohloff}, {Rousset},
  {Sanchez-Bermudez}, {Scheithauer}, {Sch{\"o}ller}, {Schuler}, {Spyromilio},
  {Straub}, {Straubmeier}, {Sturm}, {Tacconi}, {Tristram}, {Vincent}, {von
  Fellenberg}, {Wank}, {Waisberg}, {Widmann}, {Wieprecht}, {Wiest},
  {Wiezorrek}, {Woillez}, {Yazici}, {Ziegler}, \& {Zins}}]{2018A&A...615L..15G}
{GRAVITY Collaboration}, {Abuter}, R., {Amorim}, A., {et~al.} 2018, \aap, 615,
  L15

\bibitem[{Grillmair \& Dionatos(2006)}]{Grillmair_2006}
Grillmair, C.~J. \& Dionatos, O. 2006, The Astrophysical Journal, 643, L17

\bibitem[{Harris {et~al.}(2020)Harris, Millman, van~der Walt, Gommers,
  Virtanen, Cournapeau, Wieser, Taylor, Berg, Smith, Kern, Picus, Hoyer, van
  Kerkwijk, Brett, Haldane, del R{\'{i}}o, Wiebe, Peterson,
  G{\'{e}}rard-Marchant, Sheppard, Reddy, Weckesser, Abbasi, Gohlke, \&
  Oliphant}]{harris2020array}
Harris, C.~R., Millman, K.~J., van~der Walt, S.~J., {et~al.} 2020, Nature, 585,
  357

\bibitem[{{Helmi}(2020)}]{2020ARA&A..58..205H}
{Helmi}, A. 2020, \araa, 58, 205

\bibitem[{{Helmi} {et~al.}(1999){Helmi}, {White}, {de Zeeuw}, \&
  {Zhao}}]{1999Natur.402...53H}
{Helmi}, A., {White}, S. D.~M., {de Zeeuw}, P.~T., \& {Zhao}, H. 1999, \nat,
  402, 53

\bibitem[{Honma {et~al.}(2012)Honma, Nagayama, Ando, Bushimata, Choi, Handa,
  Hirota, Imai, Jike, Kim, Kameya, Kawaguchi, Kobayashi, Kurayama, Kuji,
  Matsumoto, Manabe, Miyaji, Motogi, Nakagawa, Nakanishi, Niinuma, Oh, Omodaka,
  Oyama, Sakai, Sato, Sato, Shibata, Shiozaki, Sunada, Tamura, Ueno, \&
  Yamauchi}]{Honma_2012}
Honma, M., Nagayama, T., Ando, K., {et~al.} 2012, Publications of the
  Astronomical Society of Japan, 64

\bibitem[{Hunter(2007)}]{Hunter:2007}
Hunter, J.~D. 2007, Computing in Science \& Engineering, 9, 90

\bibitem[{Ibata {et~al.}(2020)Ibata, Thomas, Famaey, Malhan, Martin, \&
  Monari}]{Ibata_2020}
Ibata, R., Thomas, G., Famaey, B., {et~al.} 2020, The Astrophysical Journal,
  891, 161

\bibitem[{Ibata {et~al.}(2016)Ibata, Lewis, \& Martin}]{Ibata_2016}
Ibata, R.~A., Lewis, G.~F., \& Martin, N.~F. 2016, The Astrophysical Journal,
  819, 1

\bibitem[{{Ibata} {et~al.}(2017){Ibata}, {Lewis}, {Thomas}, {Martin}, \&
  {Chapman}}]{2017ApJ...842..120I}
{Ibata}, R.~A., {Lewis}, G.~F., {Thomas}, G., {Martin}, N.~F., \& {Chapman}, S.
  2017, \apj, 842, 120

\bibitem[{{Jiao} {et~al.}(2023){Jiao}, {Hammer}, {Wang}, {Wang}, {Amram},
  {Chemin}, \& {Yang}}]{Jiao2023}
{Jiao}, Y., {Hammer}, F., {Wang}, H., {et~al.} 2023, arXiv e-prints,
  arXiv:2309.00048

\bibitem[{Johnson(2007)}]{NLopt}
Johnson, S.~G. 2007, The {NLopt} nonlinear-optimization package,
  \url{https://github.com/stevengj/nlopt}

\bibitem[{{Johnston} \& {Carlberg}(2016)}]{2016ASSL..420..169J}
{Johnston}, K.~V. \& {Carlberg}, R.~G. 2016, in Astrophysics and Space Science
  Library, Vol. 420, Tidal Streams in the Local Group and Beyond, ed. H.~J.
  {Newberg} \& J.~L. {Carlin}, 169

\bibitem[{{Johnston} {et~al.}(1999){Johnston}, {Zhao}, {Spergel}, \&
  {Hernquist}}]{1999ApJ...512L.109J}
{Johnston}, K.~V., {Zhao}, H., {Spergel}, D.~N., \& {Hernquist}, L. 1999,
  \apjl, 512, L109

\bibitem[{Klein(1949)}]{RevModPhys.21.531}
Klein, O. 1949, Rev. Mod. Phys., 21, 531

\bibitem[{Koposov {et~al.}(2010)Koposov, Rix, \& Hogg}]{Koposov_2010}
Koposov, S.~E., Rix, H.-W., \& Hogg, D.~W. 2010, The Astrophysical Journal,
  712, 260

\bibitem[{{Koposov} {et~al.}(2010){Koposov}, {Rix}, \&
  {Hogg}}]{2010ApJ...712..260K}
{Koposov}, S.~E., {Rix}, H.-W., \& {Hogg}, D.~W. 2010, \apj, 712, 260

\bibitem[{{Krut} {et~al.}(2023){Krut}, {Arg{\"u}elles}, {Chavanis}, {Rueda}, \&
  {Ruffini}}]{2023ApJ...945....1K}
{Krut}, A., {Arg{\"u}elles}, C.~R., {Chavanis}, P.~H., {Rueda}, J.~A., \&
  {Ruffini}, R. 2023, \apj, 945, 1

\bibitem[{{Law} {et~al.}(2009){Law}, {Majewski}, \&
  {Johnston}}]{2009ApJ...703L..67L}
{Law}, D.~R., {Majewski}, S.~R., \& {Johnston}, K.~V. 2009, \apjl, 703, L67

\bibitem[{{Lux} {et~al.}(2013){Lux}, {Read}, {Lake}, \&
  {Johnston}}]{2013MNRAS.436.2386L}
{Lux}, H., {Read}, J.~I., {Lake}, G., \& {Johnston}, K.~V. 2013, \mnras, 436,
  2386

\bibitem[{{Malhan} \& {Ibata}(2019)}]{2019MNRAS.486.2995M}
{Malhan}, K. \& {Ibata}, R.~A. 2019, \mnras, 486, 2995

\bibitem[{{Malhan} {et~al.}(2019){Malhan}, {Ibata}, {Carlberg}, {Valluri}, \&
  {Freese}}]{2019ApJ...881..106M}
{Malhan}, K., {Ibata}, R.~A., {Carlberg}, R.~G., {Valluri}, M., \& {Freese}, K.
  2019, \apj, 881, 106

\bibitem[{Malhan {et~al.}(2018)Malhan, Ibata, Goldman, Martin, Magnier, \&
  Chambers}]{10.1093/mnras/sty1338}
Malhan, K., Ibata, R.~A., Goldman, B., {et~al.} 2018, Monthly Notices of the
  Royal Astronomical Society, 478, 3862

\bibitem[{{Malhan} {et~al.}(2021){Malhan}, {Valluri}, \&
  {Freese}}]{2021MNRAS.501..179M}
{Malhan}, K., {Valluri}, M., \& {Freese}, K. 2021, \mnras, 501, 179

\bibitem[{Martínez-Delgado {et~al.}(2010)Martínez-Delgado, Gabany, Crawford,
  Zibetti, Majewski, Rix, Fliri, Carballo-Bello, Bardalez-Gagliuffi,
  Peñarrubia, Chonis, Madore, Trujillo, Schirmer, \&
  McDavid}]{Mart_Delgado_2010}
Martínez-Delgado, D., Gabany, R.~J., Crawford, K., {et~al.} 2010, The
  Astronomical Journal, 140, 962

\bibitem[{{Merafina} \& {Ruffini}(1989)}]{1989A&A...221....4M}
{Merafina}, M. \& {Ruffini}, R. 1989, \aap, 221, 4

\bibitem[{{Mestre} {et~al.}(2020){Mestre}, {Llinares}, \&
  {Carpintero}}]{2020MNRAS.492.4398M}
{Mestre}, M., {Llinares}, C., \& {Carpintero}, D.~D. 2020, \mnras, 492, 4398

\bibitem[{Nelder \& Mead(1965)}]{NELDERMEAD}
Nelder, J.~A. \& Mead, R. 1965, The Computer Journal, 7, 308

\bibitem[{{Nibauer} {et~al.}(2022){Nibauer}, {Belokurov}, {Cranmer}, {Goodman},
  \& {Ho}}]{2022ApJ...940...22N}
{Nibauer}, J., {Belokurov}, V., {Cranmer}, M., {Goodman}, J., \& {Ho}, S. 2022,
  \apj, 940, 22

\bibitem[{{Ou} {et~al.}(2023){Ou}, {Eilers}, {Necib}, \& {Frebel}}]{Ou2023}
{Ou}, X., {Eilers}, A.-C., {Necib}, L., \& {Frebel}, A. 2023, arXiv e-prints,
  arXiv:2303.12838

\bibitem[{{Ou} {et~al.}(2024){Ou}, {Eilers}, {Necib}, \&
  {Frebel}}]{Ou_2024MNRAS}
{Ou}, X., {Eilers}, A.-C., {Necib}, L., \& {Frebel}, A. 2024, \mnras, 528, 693

\bibitem[{{Palau} \& {Miralda-Escud{\'e}}(2023)}]{2023MNRAS.524.2124P}
{Palau}, C.~G. \& {Miralda-Escud{\'e}}, J. 2023, \mnras, 524, 2124

\bibitem[{{Pouliasis} {et~al.}(2017){Pouliasis}, {Di Matteo}, \&
  {Haywood}}]{2017A&A...598A..66P}
{Pouliasis}, E., {Di Matteo}, P., \& {Haywood}, M. 2017, \aap, 598, A66

\bibitem[{Price-Whelan {et~al.}(2020)Price-Whelan, Sipőcz, Lenz, Greco,
  Starkman, Foreman-Mackey, Lim, Oh, Koposov, \&
  Major}]{adrian_price_whelan_2020_4159870}
Price-Whelan, A., Sipőcz, B., Lenz, D., {et~al.} 2020, adrn/gala: v1.3

\bibitem[{Price-Whelan(2017)}]{gala}
Price-Whelan, A.~M. 2017, The Journal of Open Source Software, 2

\bibitem[{Price-Whelan \& Bonaca(2018)}]{Price-Whelan_2018}
Price-Whelan, A.~M. \& Bonaca, A. 2018, The Astrophysical Journal Letters, 863,
  L20

\bibitem[{{Price-Whelan} {et~al.}(2016){Price-Whelan}, {Johnston}, {Valluri},
  {Pearson}, {K{\"u}pper}, \& {Hogg}}]{2016MNRAS.455.1079P}
{Price-Whelan}, A.~M., {Johnston}, K.~V., {Valluri}, M., {et~al.} 2016, \mnras,
  455, 1079

\bibitem[{{Qian} {et~al.}(2022){Qian}, {Arshad}, \&
  {Bovy}}]{2022MNRAS.511.2339Q}
{Qian}, Y., {Arshad}, Y., \& {Bovy}, J. 2022, \mnras, 511, 2339

\bibitem[{{Ramos} {et~al.}(2022){Ramos}, {Antoja}, {Yuan}, {Arentsen}, {Oria},
  {Famaey}, {Ibata}, {Mateu}, \& {Carballo-Bello}}]{2022A&A...666A..64R}
{Ramos}, P., {Antoja}, T., {Yuan}, Z., {et~al.} 2022, \aap, 666, A64

\bibitem[{{Reino} {et~al.}(2021){Reino}, {Rossi}, {Sanderson}, {Sellentin},
  {Helmi}, {Koppelman}, \& {Sharma}}]{2021MNRAS.502.4170R}
{Reino}, S., {Rossi}, E.~M., {Sanderson}, R.~E., {et~al.} 2021, \mnras, 502,
  4170

\bibitem[{{Ruffini} {et~al.}(2015){Ruffini}, {Arg{\"u}elles}, \&
  {Rueda}}]{2015MNRAS.451..622R}
{Ruffini}, R., {Arg{\"u}elles}, C.~R., \& {Rueda}, J.~A. 2015, \mnras, 451, 622

\bibitem[{{Salucci, P.} {et~al.}(2010){Salucci, P.}, {Nesti, F.}, {Gentile,
  G.}, \& {Frigerio Martins, C.}}]{Salucci2010}
{Salucci, P.}, {Nesti, F.}, {Gentile, G.}, \& {Frigerio Martins, C.} 2010,
  A\&A, 523, A83

\bibitem[{{Schive} {et~al.}(2014){Schive}, {Chiueh}, \&
  {Broadhurst}}]{2014NatPh..10..496S}
{Schive}, H.-Y., {Chiueh}, T., \& {Broadhurst}, T. 2014, Nature Physics, 10,
  496

\bibitem[{Schönrich {et~al.}(2010)Schönrich, Binney, \& Dehnen}]{Shonrich}
Schönrich, R., Binney, J., \& Dehnen, W. 2010, Monthly Notices of the Royal
  Astronomical Society, 403, 1829

\bibitem[{Shere(1974)}]{shere}
Shere, K. 1974, Commun. ACM, 17, 471

\bibitem[{Sofue(2013)}]{sofue_rotation_2013}
Sofue, Y. 2013, Publications of the Astronomical Society of Japan, 65, 118

\bibitem[{Sofue(2020)}]{sofue_2020}
Sofue, Y. 2020, Galaxies, 8

\bibitem[{{Thomas} {et~al.}(2017){Thomas}, {Famaey}, {Ibata}, {L{\"u}ghausen},
  \& {Kroupa}}]{2017A&A...603A..65T}
{Thomas}, G.~F., {Famaey}, B., {Ibata}, R., {L{\"u}ghausen}, F., \& {Kroupa},
  P. 2017, \aap, 603, A65

\bibitem[{Tolman(1930)}]{PhysRev.35.904}
Tolman, R.~C. 1930, Phys. Rev., 35, 904

\bibitem[{Van~Rossum \& Drake~Jr(1995)}]{van1995python}
Van~Rossum, G. \& Drake~Jr, F.~L. 1995, Python tutorial (Centrum voor Wiskunde
  en Informatica Amsterdam, The Netherlands)

\bibitem[{{Vera-Ciro} \& {Helmi}(2013)}]{2013ApJ...773L...4V}
{Vera-Ciro}, C. \& {Helmi}, A. 2013, \apjl, 773, L4

\bibitem[{Virtanen {et~al.}(2020)Virtanen, Gommers, Oliphant, Haberland, Reddy,
  Cournapeau, Burovski, Peterson, Weckesser, Bright, {van der Walt}, Brett,
  Wilson, Millman, Mayorov, Nelson, Jones, Kern, Larson, Carey, Polat, Feng,
  Moore, {VanderPlas}, Laxalde, Perktold, Cimrman, Henriksen, Quintero, Harris,
  Archibald, Ribeiro, Pedregosa, {van Mulbregt}, \& {SciPy 1.0
  Contributors}}]{2020SciPy-NMeth}
Virtanen, P., Gommers, R., Oliphant, T.~E., {et~al.} 2020, Nature Methods, 17,
  261

\bibitem[{{Watkins} {et~al.}(2010){Watkins}, {Evans}, \&
  {An}}]{2010MNRAS.406..264W}
{Watkins}, L.~L., {Evans}, N.~W., \& {An}, J.~H. 2010, \mnras, 406, 264

\bibitem[{{Zhao} {et~al.}(1999){Zhao}, {Johnston}, {Hernquist}, \&
  {Spergel}}]{1999A&A...348L..49Z}
{Zhao}, H., {Johnston}, K.~V., {Hernquist}, L., \& {Spergel}, D.~N. 1999, \aap,
  348, L49

\end{thebibliography}

\begin{appendix}
\section{Numerical solution of the system of differential equations}
\label{app:numerical}
In this section we will explain how we solved numerically the physical equations defined in Sec.~\ref{sec:fermionicDM}.
We start defining some constants:
\begin{align}
    \rho_{\bullet}&= g\pi^{3/2} m^4 c^3 h^{-3},\\
    r_\bullet &= c/(8\pi G \rho_\bullet)^{1/2},
\end{align}
and introduce the following changes of variables:
\begin{align}
   \zeta &= \ln(r/r_\bullet),\\
   z(\zeta) &=\ln\psi(r(\zeta)),\\
   \tilde{\nu}(\zeta) &=\nu(r(\zeta)),\\
   \beta(\zeta)&=\frac{kT(r(\zeta))}{mc^2},\\
   \alpha(\zeta)&=\frac{\mu(r(\zeta))}{mc^2},\\
   \epsilon_c(\zeta)&=\frac{E_c(r(\zeta))}{mc^2},\\
   \epsilon(p)&=\frac{E(p)}{mc^2},
\end{align}
where
\begin{equation}
    \psi(r) = 1-\textrm{e}^{-\lambda(r)} = \frac{2G}{c^2}\frac{M(r)}{r},
    \label{mass_dm}
\end{equation}
into equations~(\ref{mass_def}), (\ref{tov}) and~(\ref{tke}), obtaining:
\begin{align}
   \label{sode}
   \frac{\mathrm{d}z}{\mathrm{d}\zeta} & = -1+\mathrm{e}^{2\zeta-z}\frac{\tilde{\rho}(\zeta)}{\rho_{\bullet}},\\
   \label{sode2}
   \frac{\mathrm{d}\tilde{\nu}}{\mathrm{d}\zeta} & = \left(\mathrm{e}^{z}+\mathrm{e}^{2\zeta}\frac{\tilde{P}(\zeta)}{\rho_{\bullet}c^2}\right)(1-\mathrm{e}^{z})^{-1},\\
   \frac{1}{\beta}\frac{\mathrm{d}\beta}{\mathrm{d}\zeta}&=\frac{1}{\alpha}\frac{\mathrm{d}\alpha}{\mathrm{d}\zeta}=
   \frac{1}{\epsilon_\mathrm{c}}\frac{\mathrm{d} \epsilon_\mathrm{c}}{\mathrm{d}\zeta}=-\frac{1}{2}\frac{\mathrm{d}\tilde{\nu}}{\mathrm{d}\zeta}.\label{tke2}
\end{align}

The thermodynamical quantities, density and pressure, are given by:

\begin{align}
    \tilde{\rho}(\zeta)&=\frac{4\rho_{\bullet}}{\sqrt{\pi}}\int^\infty_1\epsilon^2[\epsilon^2-1]^{1/2}\tilde{f}(\zeta,\epsilon)\mathrm{d}\epsilon,\\
   \tilde{P}(\zeta)&=\frac{4c^2\rho_{\bullet}}{3\sqrt{\pi}}\int^\infty_1[\epsilon^2-1]^{3/2}\tilde{f}(\zeta,\epsilon)\mathrm{d}\epsilon,
\end{align}
where the fermionic King distribution as a function of
$\epsilon=E/mc^2$, in the new variables is given by:
\begin{equation}
\tilde{f}(\zeta,\epsilon)\equiv \frac{h^3}{g}f(r(\zeta),p(\epsilon))=
      \frac{1-\mathrm {e}^{[\epsilon-\epsilon_\mathrm{c}(\zeta)]/\beta(\zeta)}}
      {1+\mathrm {e}^{[\epsilon-\alpha(\zeta)]/\beta(\zeta)}}\quad\mathrm{if}\quad \epsilon \leq \epsilon_\mathrm{c}(\zeta),
\end{equation}
and $\tilde{f}(\zeta,\epsilon) = 0$ otherwise.

Note that equations~(\ref{tke2}) can be analytically integrated to obtain:
\begin{align}
 \beta(\zeta) &= \beta_0\mathrm{e}^{-\tilde{\nu}(\zeta)/2}, \nonumber\\
 \alpha(\zeta) &= \alpha_0\mathrm{e}^{-\tilde{\nu}(\zeta)/2}, \nonumber\\
 \epsilon_c(\zeta) &= \epsilon_{c0}\mathrm{e}^{-\tilde{\nu}(\zeta)/2},
 \label{tke_solved}
\end{align}
thus transforming the original system of five differential equations, i.e.~(\ref{sode}-\ref{tke2}), to a system of just two differential equations to be solved numerically subject to the constraints (\ref{tke_solved}).

It is not possible to integrate these equations from $r=0$ because $\zeta(r)$ diverges at the origin.
Therefore, the following approximations for the initial conditions at a value $r_{\rm{min}}\gtrsim 0$ were used:
\begin{align}
   \nu(r_{\rm{min}}) &= \frac{1}{3}\frac{\rho_0}{\rho_\bullet}\left[\frac{r_{\rm{min}}}{r_b}\right]^2\equiv \tau, \\
   \psi(r_{\rm{min}})&= \frac{1}{3}\frac{\rho_0}{\rho_\bullet}\left[\frac{r_{\rm{min}}}{r_\bullet}\right]^2 = \tau,
\end{align}
which implies
\begin{equation}
   \frac{r_{\rm{min}}}{r_\bullet}=\sqrt{3\tau\frac{\rho_\bullet}{\rho_0}},
\end{equation}
where
$\tau\equiv 2\times10^{-15}$ and

\begin{equation}
    \rho_0\equiv \rho(0) = \frac{4\rho_{\bullet}}{\sqrt{\pi}}\int^\infty_1\epsilon^2[\epsilon^2-1]^{1/2}\tilde{f}_0(\epsilon)\mathrm{d}\epsilon,
\end{equation}

where

\begin{equation}
\tilde{f}_0(\epsilon)=
      \frac{1-\mathrm {e}^{[\epsilon-\epsilon_\mathrm{c0}]/\beta_0}}
      {1+\mathrm {e}^{[\epsilon-\alpha_0]/\beta_0}}\quad\mathrm{if}\quad \epsilon \leq \epsilon_\mathrm{c0},
\end{equation}
and $\tilde{f}_0(\epsilon)=0$ otherwise.

In this way, the initial conditions of our numerical system is given
by $\zeta_{\rm{min}}=\zeta(r_{\rm{min}})$,
$\tilde{\nu}_{\rm{min}}=\tau$ and $z_{\rm{min}}=\ln{\tau}$, and the system parameters to be varied are $m$, $\beta_0$, $\alpha_0$ and $\epsilon_{c0}$.
In agreement with Eqs.~(\ref{rar_params}), we will use as parameters the following normalized quantities: $m$, $\beta_0$, $\theta_0=(\alpha_0-1)/\beta_0$, and $W_0=(\epsilon_{\mathrm{c}0}-1)/\beta_0$, or $\omega_0=W_0-\theta_0$ instead of $W_0$ in some cases.

Equations~(\ref{sode}) and (\ref{sode2}) were solved using
a {\sc Python}~\citep{van1995python} script\footnote{\texttt{model\_def.py}}
that makes use of the {\it NumPy}~\citep{2020SciPy-NMeth} and {\it SciPy}~\citep{harris2020array} libraries, under the \texttt{LSODA} algorithm as solver.  We used relative and absolute tolerance parameters given by \texttt{rtol}$=5\times10^{-14}$ and \texttt{atol}$=0$, respectively.

After obtaining the numerical solution, since the right hand sides of Eqns.~(\ref{sode}) and (\ref{sode2}) do not depend on the metric but only on its radial derivative, we can add
a constant $\tilde{\nu}_0$ to the solution in order to satisfy a condition of continuity with the Schwarzschild metric at the border of the fermion distribution, obtaining:
\begin{equation}
\tilde{\nu}_0 = 2\ln\left(\frac{\beta_\mathrm{b}}{\beta_0}\sqrt{1-\psi_\mathrm{b}}\right),
\end{equation}
where $\psi_\mathrm{b}$ and $\beta_\mathrm{b}$ are quantities evaluated at the border, that is when
$\rho(r_\mathrm{b})\rightarrow 0$.\footnote{The border was defined numerically as the radius in which
the density decays to $\rho_b=10^{-10}~M_\odot~\rm{pc}^{-3}$.}

\end{appendix}

\end{document}